\documentclass[10pt,journal,twoside,final]{IEEEtran}
\UseRawInputEncoding

\IEEEoverridecommandlockouts

\makeatletter
\newcommand{\biggg}[1]{{\hbox{$\left#1\vbox to 20.5pt{}\right.\n@space$}}}
\newcommand{\Biggg}[1]{{\hbox{$\left#1\vbox to 23.5pt{}\right.\n@space$}}}
\newcommand{\bigggg}[1]{{\hbox{$\left#1\vbox to 26.5pt{}\right.\n@space$}}}
\newcommand{\Bigggg}[1]{{\hbox{$\left#1\vbox to 29.5pt{}\right.\n@space$}}}
\newcommand{\biggggg}[1]{{\hbox{$\left#1\vbox to 32.5pt{}\right.\n@space$}}}
\newcommand{\Biggggg}[1]{{\hbox{$\left#1\vbox to 35.5pt{}\right.\n@space$}}}
\newcommand{\bigggggg}[1]{{\hbox{$\left#1\vbox to 38.5pt{}\right.\n@space$}}}
\newcommand{\Bigggggg}[1]{{\hbox{$\left#1\vbox to 41.5pt{}\right.\n@space$}}}
\makeatother

\usepackage{bm,cite,algorithm,algorithmicx,float,amsmath,amssymb}
\usepackage[dvips]{graphicx}
\usepackage{subfigure,amsfonts}
\usepackage{mdwmath}
\usepackage{mdwtab}
\usepackage{xcolor}
\usepackage{balance}
\floatname{algorithm}{Algorithm}
\usepackage{titlesec}
\usepackage{multirow}
\allowdisplaybreaks
\definecolor{myGreen}{rgb}{0.76, 0.93, 0.63}
\definecolor{myGreen2}{rgb}{0.92, 0.92, 0.92}

\begin{document}


\title{Dual-UAV-Aided Covert Communications\\ for Air-to-Ground ISAC Networks}

\author{Jingke Sun, 
Liang Yang, \IEEEmembership{Senior Member,~IEEE,}
Alexandros-Apostolos A. Boulogeorgos, \IEEEmembership{Senior Member,~IEEE,}
Theodoros A. Tsiftsis, \IEEEmembership{Senior Member,~IEEE,}
and Hongwu~Liu, \IEEEmembership{Senior Member,~IEEE}
\thanks{J. Sun and H. Liu are with the School of Information Science and Electrical Engineering, Shandong Jiaotong University, Jinan 250357, China (e-mail: 205029@sdjtu.edu.cn,  liuhongwu@sdjtu.edu.cn).}
\thanks{L. Yang is with the College of Computer Science and Electronic Engineering, Hunan University, Changsha 410082, China (e-mail: liangy@hnu.edu.cn).}
\thanks{A. A. Boulogeorgos is with the Department of Electrical and Computer Engineering, University of Western Macedonia, 50100 Kozani, Greece (e-mail: aboulogeorgos@uowm.gr).}
\thanks{T. A. Tsiftsis is with the Department of Informatics and Telecommunications, University of Thessaly, 35100 Lamia, Greece (e-mail: tsiftsis@uth.gr).}
}
\maketitle

\setcounter{page}{1}
\begin{abstract}
To enhance both the sensing and covert communication performance, a dual-unmanned aerial vehicle (UAV)-aided scheme is proposed for integrated sensing and communication networks, in which one UAV maneuvers as the aerial dual-functional base-station (BS), while another UAV flies as the cooperative jammer. Artificial noise (AN) transmitted by the jamming UAV is utilized not only to confuse the ground warden but also to aid the aerial BS to sense multiple ground targets by combing the target-echoed dual-functional waveform and AN components from a perspective of the hybrid monostatitc-bistatic radar. We employ the distance-normalized beampattern sum-gain to measure the sensing performance. To maximize the average covert rate (ACR) from the aerial BS to the ground user,  the dual-functional BS beamforming, jamming UAV beamforming, and dual-UAV trajectory are co-designed, subject to transmit power budgets, UAV maneuver constraint, covertness requirement, and sensing performance constraint.  The imperfect successive interference cancellation (SIC) effects on the received signal-to-interference-plus-noise ratio are also considered in maximizing the ACR. To tackle the highly complicated non-convex ACR maximization problem, dual-UAV beamforming and dual-UAV trajectory are optimized in a block coordinate descent way using the trust-region successive convex approximation and semidefinite relaxation. To find the dual-UAV maneuver locations suitable for sensing the ground targets, we first optimize the dual-UAV trajectory for the covert communication purpose only and then solve a weighted distance minimization problem for the covert communication and sensing purpose. Simulations highlight the superior ACR and sensing performance of the proposed dual-UAV-aided covert communication scheme and reveal the effects of imperfect SIC on the system performance. The benefits of deploying the jamming UAV to enhance both the ACR and sensing performance are also quantified.   
\end{abstract}


\section{Introduction}

By sharing hardware resources and spectrum between sensing and communication functions, integrated sensing and communication (ISAC) enhances spectral efficiency while reducing infrastructure costs \cite{DUAL_function_ISAC, ISAC_Resource_Framework, enabling_Joint_communication}. The convergence of sensing and communication brought by ISAC has been regarded as a paradigm shift in wireless networks, creating intelligent applications that dynamically adapt to user needs and physical surroundings \cite{enabling_Joint_communication}. 
Due to advantages in flexible deployment, dynamic coverage enhancement, and improved line-of-sight (LoS) connectivity, drones, including unmanned aerial vehicles (UAVs) and autonomous aerial vehicles (AAV), have garnered significant attention from both the industrial and academic worlds as promising platforms for emerging air-to-ground (A2G) ISAC systems \cite{UAV_meets_ISAC, UAV_Periodic_ISAC, cooperative_trajectory_ISAC, Semantic_AAV_ISAC}. A2G-ISAC aided by UAV and AAV demonstrates particular efficacy in enabling real-time environmental monitoring, precision target tracking, and reliable high-throughput data transmission across diverse application scenarios, including disaster response operations, smart city infrastructures, and large-scale ecological surveillance initiatives \cite{UAV_Periodic_ISAC_Throughput, UAV_Adaptable_ISAC, Joint_Maneuver, Joint_radar_ISAC}.

The open nature of wireless medium exposes ISAC systems to spoofing, jamming, and denial-of-service attacks, while the dual-functionality of ISAC amplifies risks by merging sensing and data transmission pathways \cite{ISAC_6G_Sucure}. In particular, A2G-ISAC systems face critical security challenges, including vulnerabilities of drones to hijacking or destruction, eavesdropping and data manipulation in wireless links, and privacy breaches due to unauthorized sensing of sensitive information. Additionaly, limited onboard computational resources constrain robust encryption implementation, and dynamic network topologies complicate secure authentication and trust management between UAVs, users, and ground infrastructure.
Using inherent channel characteristics to prevent eavesdropping, physical layer security (PLS) has been proposed as an effective approach to secure confidential data transmission in A2G-ISAC systems
\cite{PLS_ISAC_AAV, Secrecy_AAV_ISAC, UAVs_meet_ISAC_Secure}. In \cite{PLS_ISAC_AAV}, the  AAV trajectory and transceiver beamformers were jointly optimized to maximize the average secrecy rate, while the A2G sensing accuracy with respect to an untrusted target was guaranteed. An AAV was deployed as a multi-functional aerial base-station (BS) to serve ground users, meanwhile detecting and jamming a ground eavesdropper \cite{Secrecy_AAV_ISAC}.    

To ensure secure and reliable transmissions of sensitive data while maintaining operational stealth for A2G-ISAC systems, it is imperative to prevent eavesdropping and adversarial detection in strong LoS dominated environments, whereas PLS techniques primarily focus on resisting eavesdropping without addressing the existence of transmission behaviors. Recently, covert communications have been employed for A2G-ISAC systems to mitigate interception and unauthorized detection during sensitive missions, ensuring secure and reliable integrated sensing and communication operations \cite{UAV_Covert_ISAC_Xingwang, UAV_ISAC_Covert}. 
Specifically, in an A2G-ISAC system consisting of multiple ground covert users and multiple ground wardens, the UAV trajectory and beamforming were iteratively optimized by a block coordinate descent (BCD) algorithm to improve average covert rate (ACR) and ensure sensing capability regarding multiple targets located within a specified sensing area \cite{UAV_Covert_ISAC_Xingwang}. To  maximize the minimum achievable covert rate among all ground users for an A2G-ISAC system, the UAV trajectory, transmit power allocation, and communication scheduling were iteratively optimized  \cite{UAV_ISAC_Covert}.  

However, the aforementioned research on covert communications in A2G-ISAC systems has predominantly focused on single-UAV scenarios, overlooking the potential advantages of collaborative multi-UAV deployments. Cooperative strategies among multiple UAVs could significantly enhance system performance by optimizing resource allocation, improving coverage, and strengthening interference cancellation capabilities in A2G-ISAC tasks \cite{Multi_UAV_ISAC_GCWkshps, Multi_UAV_ISAC_WCSP, Multi_UAV_ISAC_IoTJ}. In \cite{Multi_UAV_ISAC_GCWkshps} and \cite{Multi_UAV_ISAC_WCSP}, multiple UAVs simultaneously transmitted individual messages to communicate with their respective ground users and cooperatively used the information signals to detect a target on the ground. Likewise, the multi-UAV trajectory, user association, and beamforming were jointly optimized to maximize the sum weighted rate of all ground users while ensuring the sensing beampattern gain of the target \cite{Multi_UAV_ISAC_IoTJ}. On the other hand, cooperative jamming is employed in covert communications to enhance system performance by generating artificial noise (AN) or interference that masks legitimate transmissions from potential detections with low detection probabilities \cite{Covert_AN, Covert_Distributed_Jamming, Covert_UAV_Jammer, Cooperative_Jamming, Covert_Multiantenna_UAV_Jammer}. With a jamming UAV located stationarily, the user scheduling, transmit power, and source UAV trajectory were alternatively and iteratively optimized to maximize the secrecy rate for an A2G-ISAC system \cite{ISAC_UAV_Multi_Eve}. Nowadays, covert communication schemes in A2G-ISAC systems rely on single-UAV operations without integrating multi-UAV collaboration, limiting resource optimization and interference management. In addition, jamming strategies have been scarcely explored in A2G-ISAC systems, leaving untapped potential to improve covertness through cooperative UAV jamming.

 Motivated by the aforementioned discussions, we propose a dual-UAV-aided covert communication scheme for an A2G-ISAC network to enhance covertness and sensing performance. Specifically, we propose to deploy a jamming UAV to aid covert communications from a source UAV to a ground user, while the sensing of multiple ground targets are conducted by the source and jamming UAVs in a cooperative manner. In contrast to the publicly available schemes, in the proposed scheme,  the source UAV and jamming UAV formulate a hybrid monostatic-bistatic radar model to enhance both the sensing and covert communication performance. The main contributions of this work are summarized as follows:

\begin{itemize}
\item  We propose a dual-UAV-aided covert communication scheme to enhance both the 
sensing and covert communication performance. The source and jamming UAVs not only collaborate with each other to improve covertness but also construct a hybrid monostatic-bistatic radar model by combing the target-echoed dual-functional waveform transmitted by the source UAV and AN transmitted by the jamming UAV. 
\item  For the considered A2G-ISAC network, the distance-normalized beampattern sum-gain is adopted to measure the sensing performance from a perspective of the hybrid monostatitc-bistatic radar. In addition, taking into account the imperfect successive interference cancellation (SIC),  the residual interference effects on the received signal-to-interference-plus-noise ratios (SINRs) are investigated to evaluate the practical system performance. 
\item  To tackle the highly complicated non-convex ACR maximization problem, we present a BCD algorithm to optimize the dual-UAV beamforming and dual-UAV trajectory iteratively, using the trust-region successive convex approximation (SCA) and semidefinite relaxation (SDR) techniques. To determine the dual-UAV maneuver locations suitable for sensing, we first optimize the dual-UAV trajectory for the covert communication purpose only and then solve a weighted distance minimization problem for the covert communication and sensing purpose, using a heuristic greedy algorithm.
\item Simulations are conducted to quantify the covertness and sensing performance of the proposed scheme assuming different design parameters. The simulation results reveal that the proposed scheme achieves superior ACR and sensing performance for the considered A2G-ISAC network. The effects of the cooperative UAV jamming and imperfect SIC on system performance are also verified. 
\end{itemize}   

 The rest of this paper is organized as follows: in Section II,  the system model is presented including the signal model, sensing performance metric, covertness requirement, and optimization problem formulation. Section III presents the covert beamfomring and trajectory optimization for the covert communication purpose only. Section IV introduces the solution for the covert communication and sensing purpose. Simulation results are presented in Section V. Finally, conclusions are drawn in Section VI.

$Notations$: Boldface letters refer to vectors (lower case) or matrices (upper case). 
$\mathbf{A}\succeq\mathbf{0}$ means that A is positive semidefinite. $\mathbf{0}$ denotes an all-zero vector or matrix. $\mathbf{I}$ denotes an identity matrix. 
$\rm{rank}(\mathbf{A})$ and $ \rm{tr}(\mathbf{A})$  denote the rank and trace of matrix $\mathbf{A}$, respectively. matrix $[\mathbf{A}]_{k, \ell}$ denotes the element in the $k$-th row and $\ell$-th column of matrix $\mathbf{A}$. The $(\cdot)^H$ and $(\cdot)^T$ denote the conjugate transpose and transpose, respectively. $\vert\cdot\vert$ denotes the magnitude of a complex number. $\Vert\cdot\Vert$, $\Vert\cdot\Vert_*$, and $\Vert\cdot\Vert_2$ denote Euclidean norm, nuclear norm, and spectral norm, respectively. $\mathbb{C}^{{M}\times{N}}$ denotes the space of ${M}\times{N}$ complex matrices. $\mathbb{E} (\cdot)$ denotes the statistical expectation.

\section{System Model}
\begin{center}
\begin{figure}[tb]
~~~~~\includegraphics[width=3.2in]{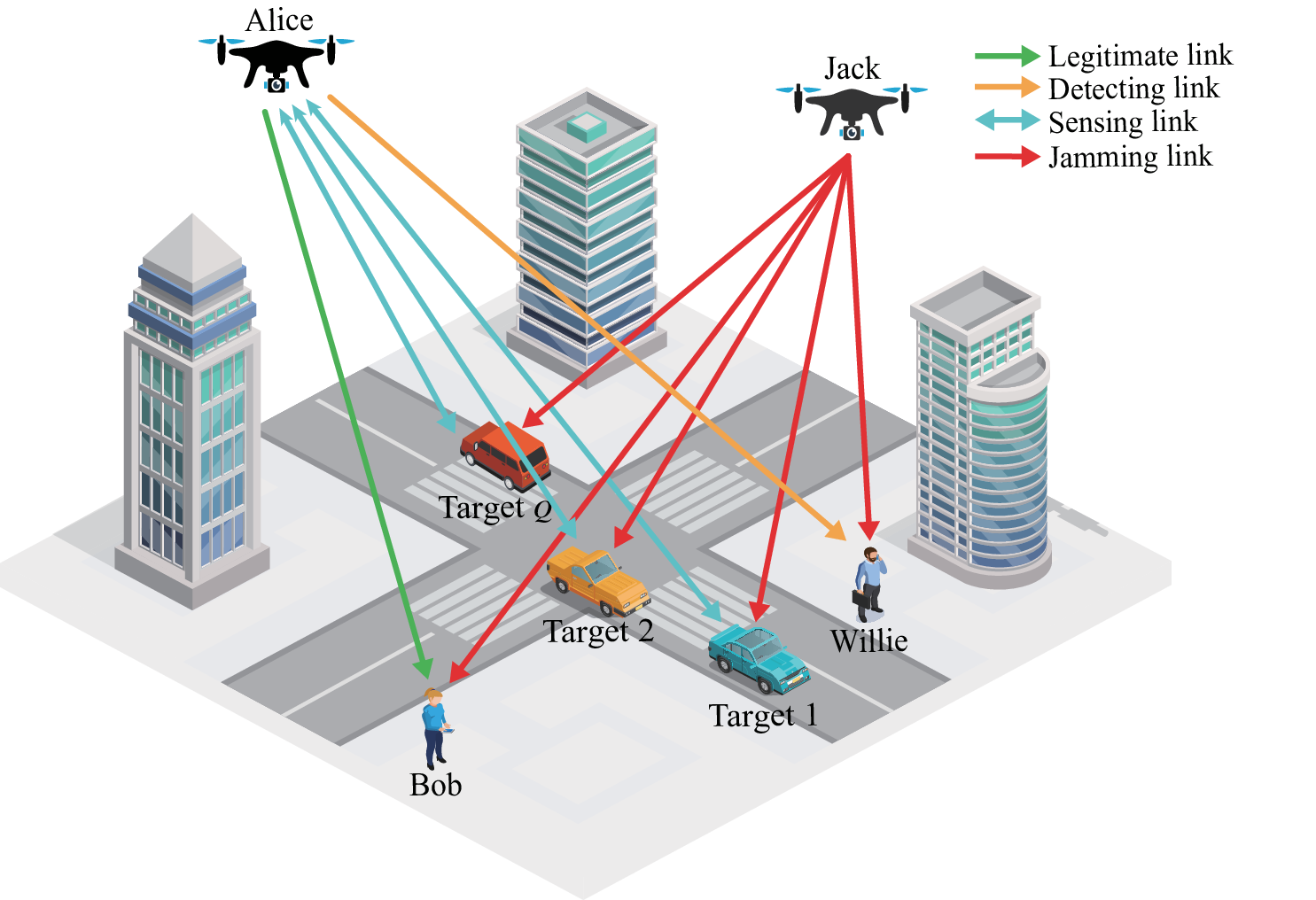}
\caption{The A2G ISAC network model}
\label{fig:system_model}
\vspace{-0.2in}
\end{figure}
\end{center}

As shown in Fig. 1, we consider covert communication in an A2G ISAC network, which consists of a dual-function UAV (Alice), a cooperative jamming UAV (Jack), a ground receiver (Bob), a ground warden (Willie), and $Q$ ground targets. Alice wishes to covertly communicate with the intended user Bob with a low probability of being detected by Willie, while sensing  the ground targets. On the other hand, Jack transmits the AN signal to confuse Willie's detection and aid Alice to sense the ground targets. In particular, Alice detects the ground targets based on the echoed ISAC wave and AN components; Thus, formulating a hybrid monostatic-bistatic radar system \cite{Hybrid_Multistatic,Hybrid_Bistatic_Monostatic}.    
We assume that Alice and Jack are both equipped with a vertically positioned uniform linear array (ULA) comprising $M$ antennas, while Bob and Willie are, respectively, equipped with a single receive antenna. 

In the considered ISAC network, a finite task period $\mathcal{T} \triangleq [0, T]$ is divided into $N$ time slots with  time slot $ n \in \mathcal{N} \triangleq \{1,2, \cdots, N\}$ having a duration $ \Delta_t = T/N $, where $T>0$ is the duration of the task period. To facilitate the joint design of the dual-UAV trajectory and beamforming, $\Delta_t$ is chosen as small as possible so that the location of each UAV within a time slot can be roughly regarded unchanged. 
Without loss of generality, we assume that both UAVs are able to transmit signals continuously and consistently for the covert communication purpose throughout the entire task period. In areas near targets, both UAVs collaborate in the covert communication as well as in the hybrid monostatic-bistatic radar sensing \cite{Hybrid_Multistatic,Hybrid_Bistatic_Monostatic}. 
Within each task period, the time slots are classified into the covert communication only (CCO)  and covert communication and sensing (CCS) phases. Specifically, in all time slots of the CCO phase $\mathcal{N}_c \triangleq \{1,2, \cdots, N_c\}$,  both UAVs only engage in the covert communication. In all time slots of the CCS phase $\mathcal{N}_s \triangleq \{1,2, \cdots, N_s\}$, both UAVs simultaneously conduct the covert communication and target sensing. In addition, the numbers of time slots in the CCO and CCS phases satisfy $1 \le N_c < N$, $Q \le N_s < N$, $N_c + N_s = N$.
In time slot $n$, the horizontal location of UAV $m$  ($m \in \{a, j \}$) is denoted as $\mathbf{u}_m[n]=\big[x_m[n],y_m[n]\big]^T$, where $m$ can be $a$ and $j$ to represent Alice and Jack, respectively. Within a task period, the predetermined initial and final locations of UAV $m$ are denoted as $\mathbf{u}_m^{\rm{I}}[0]$ and $\mathbf{u}_m^{\rm{F}}[N]$, respectively. The location of ground node $\bar m$ (${\bar m} \in \{b, w\}$) is denoted as $\mathbf{v}_{\bar m}=[x_{\bar m},y_{\bar m}]^T$, where $\bar m$ can be $b$ and $w$ representing Bob and Willie, respectively. Let $A_m$ denote a minimum altitude of UAV $m$ that guarantees safe flying, taking into account obstacle avoidance and energy savings by omitting ascending or descending \cite{UAV_Trajectory_and_Beamforming,Covert_IRS}.
In a designated task period, Alice and Jack maneuver dynamically from one position to another one, subject to the following UAV maneuvering constraints, i.e.,
\begin{equation}
    {\bf u}_m[0] = {\bf u}_m^{\rm I},  ~m \in \{a, j\} ,\label{eq:position_1}
\end{equation}
\begin{equation}
    {\bf u}_m[N] =  {\bf u}_m^{\rm F},  ~m \in \{a, j\} ,\label{eq:position_N}
\end{equation}
\begin{equation}
    \Vert \mathbf{u}_m[n]-\mathbf{u}_m[n-1]\Vert \le V_{\max}, ~m \in \{a, j\}, ~\forall n \in {\cal{N}}_c, \label{eq:displacement}
\end{equation}
where ${\bf u}_m^{\rm I} = [x^{\rm I}, y^{\rm I}]^T$ and ${\bf u}_m^{\rm F} = [x^{\rm F}, y^{\rm F}]^T$ denote the UAV's initial and final horizontal locations, respectively, and $V_{\max} = \tilde{V}_{\max} \Delta_t  $ is the maximum UAV displacement within each time slot with $\tilde{V}_{\max}$ denoting the maximum maneuver speed.
 
\subsection{Signal Model} 

In both the CCO and CCS phases, the transmit signal of Alice in time slot $n$ can be expressed as
\begin{equation}
    {\mathbf{x}}_a[n] = \left\{ {\begin{array}{*{20}{c}}
{ {\mathbf{w}_{a}}[n]s_{a}[n] },& {\forall n \in {\cal N}_c},\\
{ {\mathbf{w}_{a}}[n]s_{a}[n]+ \mathbf{x}_r[n] },&{\forall n \in {\cal N}_s},
\end{array}} \right.
 \label{eq:Xa_signal}    
\end{equation}
where  $\mathbf{w}_a[n]\in \mathbb{C}^{M\times1}$ is the transmit beamformer at Alice, $ s_a[n] \sim\mathcal{CN}(0,1) $ is the information signal intended to Bob, and $\mathbf{x}_r[n] \in \mathbb{C}^{M\times1}$ is the sensing signal with zero mean and covariance matrix $\mathbf{R}_r[n]=\mathbb{E}(\mathbf{x}_r[n]\mathbf{x}_r^H[n]) \succeq 0$. We assume that $s_a[n]$ and $\mathbf{x}_r[n]$ are independent of each other. At Alice, $\mathbf{s}_a[n]$ is transmitted using a single beam, whereas the sensing signal $\mathbf{x}_r[n]$ is transmitted across $M_s$ beams with $0\le M_s\le M$. In particular, $M_s$ beams can be generated 
by the eigenvalue decomposition of $\mathbf{R}_r[n]$ with ${\rm{rank}}({\bf{R}}_r[n])=M_s$. 
In both the CCO and CCS phases, the transmit signal of Jack in time slot $n$ can be written as
\begin{equation}
    \mathbf{x}_j[n]={\mathbf{w}_{j}}[n]{s}_{j}[n],~~ \forall n \in{\cal N},
\end{equation}
where $\mathbf{w}_j[n]\in \mathbb{C}^{M\times1}$ is the transmit beamformer at Jack and $s_j[n] \sim\mathcal{CN}(0,1) $ is the AN signal. The average transmit powers of Alice in the CCO and CCS phases can be denoted as $\mathbb{E}(\|{\mathbf{x}}_a[n]\|^2) =  \| \mathbf{w}_{a} \|^2$ and $\mathbb{E}(\|{\mathbf{x}}_a[n]\|^2) =  \| \mathbf{w}_{a} \|^2 + {\rm{trace}}(\mathbf{R}_r[n])$, respectively. The average transmit power of Jack in both the CCO and CCS phases can be denoted as $\mathbb{E}(\|\mathbf{x}_j[n]\|^2) = \| \mathbf{w}_{j} \|^2$. Then, the transmit power constraints of Alice and Jack can be expressed as
\begin{equation}
    \Vert\mathbf{w}_a [n]\Vert^2 \le P_a^{\max},\quad \forall n \in {\cal N}_c,
\end{equation}
\begin{equation}
    \Vert\mathbf{w}_a [n] \Vert^2+ {\rm{trace}}(\mathbf{R}_r[n])\le P_a^{\max},\quad\forall n \in {\cal N}_s,
\end{equation} 
\begin{equation}
    \Vert\mathbf{w}_j [n]\Vert^2 \le P_j^{\max},\quad \forall n \in{\cal N},
\end{equation}
where $P_a^{\max}$ and $P_j^{\max}$ denote the maximum transmit powers of Alice and Jack, respectively. 

In general, UAV operates at a relatively high altitude, resulting in strong LoS links between UAV and ground nodes \cite{Joint_Optimization_for_Covert,Joint_Trajectory_and_Precoding,Covert_Communication_NOMA}. Thus, we model the LoS channel between UAV $m$ ($m \in \{a, j \}$) and ground node $\bar m$ (${\bar m} \in \{b, w\}$) as
\begin{eqnarray}
    \mathbf{h}_{m \bar m}(\!\mathbf{u}_m[n],\mathbf{v}_{\bar m}\!) \! &\!\!\!\!=\!\!\!\!&\!{\sqrt{\beta d^{-2}(A_m,\mathbf{u}_m[n],\mathbf{v}_{\bar m})}\mathbf{a}(\mathbf{u}_m[n],\mathbf{v}_{\bar m}\!)}\nonumber\\
    \!\!&\!\!\!\!= \!\!\!\!&\! \sqrt{\frac{\beta}{A_m^2 \!+\! \Vert \mathbf{u}_m[n] \!-\!\mathbf{v}_{\bar m}\Vert^2}}\mathbf{a}(\mathbf{u}_m[n],\mathbf{v}_{\bar m}\!),~~~
\end{eqnarray}
where $\beta$ represents the path-loss at the reference distance $d_0=1 $ m, $d_{m \bar m}(A_m,\mathbf{u}_m[n],\mathbf{v}_{\bar m})=\sqrt{A_m^2+\Vert \mathbf{u}_m[n]-\mathbf{v}_{\bar m}\Vert^2}$ denotes the distance between UAV $m$ and ground node ${\bar m}$, and ${\mathbf{a}(\mathbf{u}_m[n],\mathbf{v}_{\bar m})}$ is the steering vector given by
\begin{eqnarray}
    \mathbf{a}(\mathbf{u}_m[n],\mathbf{v}_{\bar m})\!=\! \Big[1,{{e}^{\mathfrak{j}2\pi \frac{d_m}{\lambda }\cos \theta }}, \cdots,{{e}^{\mathfrak{j}2\pi \frac{d_m}{\lambda }(M-1)\cos \theta }} \Big]^T   \label{steering_vector_a}
\end{eqnarray}
with $d_m$ denoting the spacing between two adjacent antennas at UAV $m$, $\lambda$ denoting the wavelength of the signal carrier, and $\mathfrak{j}^2 = -1 $. In \eqref{steering_vector_a}, the angle of departure (AoD) from UAV $m$ to ground node ${\bar m}$ is given by
\begin{equation}    \theta(\mathbf{u}_m[n],\mathbf{v}_{\bar m})=\arccos\frac{A_m}{\sqrt{A_m^2+\Vert \mathbf{u}_m[n]-\mathbf{v}_{\bar m} \Vert^2} }.
\end{equation}
In the CCO and CCS phases,  the received signals at Bob can be expressed as
\begin{eqnarray}  
    y_b^{\rm {\rm{cco}}}[n] &\!\!\!\!=\!\!\!\!& \mathbf{h}_{ab}^H\mathbf{x}_a[n] + \mathbf{h}_{jb}^H\mathbf{x}_j[n] + z_b[n] \nonumber \\
    &\!\!\!\!=\!\!\!\!&  \mathbf{h}_{ab}^H {\mathbf{w}_{a}}[n]{s}_{a}[n]   \!+\!  \mathbf{h}_{jb}^H{\mathbf{w}_{j}}[n]{s}_{j}[n] \!+\!  z_b[n] ~~ \label{eq:y_b_cco} 
\end{eqnarray} 
and
\begin{eqnarray}  
    y_b^{\rm {\rm{ccs}}}[n] &\!\!\!\!\!\!=\!\!\!\!\!\!& \mathbf{h}_{ab}^H\mathbf{x}_a[n] + \mathbf{h}_{jb}^H\mathbf{x}_j[n] + z_b[n] \nonumber \\
    &\!\!\!\!\!\!=\!\!\!\!\!\!&  \mathbf{h}_{ab}^H \big({\mathbf{w}_{a}}[n]{s}_{a}[n] \!+\! \mathbf{x}_r[n]\big) \!+\!  \mathbf{h}_{jb}^H{\mathbf{w}_{j}}[n]{s}_{j}[n] \!+\!  z_b[n],~~~~~ \label{eq:y_b_ccs}
\end{eqnarray} 
respectively, where $z_b[n] \sim \mathcal{CN}(0,\sigma_b^2)$ is the additive white Gaussian noise (AWGN) at Bob. 
At Bob, before detecting  $s_a[n]$, SIC is applied to eliminate the sensing and AN signals \cite{Bi_ISAC}. Unfortunately, residual interference generally exists in detecting $s_a[n]$ due to imperfect SIC.   
Based on \eqref{eq:y_b_cco} and \eqref{eq:y_b_ccs}, the received SINRs at Bob in the CCO and CCS phases can be written as
\begin{eqnarray}    
    \gamma_b^{\rm {\rm{cco}}}[n]  &\!\!\!\!=\!\!\!\!& \frac{|\mathbf{h}_{ab}^H\mathbf{w}_a[n]|^2}{  \varpi_{jb}|\mathbf{h}_{jb}^H\mathbf{w}_j[n]|^2 + \sigma_{b}^2} \label{eq:gamma_b_c}
    \end{eqnarray}
and
\begin{equation}
    \gamma_b^{\rm {\rm{ccs}}}[n]  =\frac{|\mathbf{h}_{ab}^H\mathbf{w}_a[n]|^2}{\varpi_{rb}(\mathbf{h}_{ab}^H\mathbf{R}_r\mathbf{h}_{ab}) + \varpi_{jb}|\mathbf{h}_{jb}^H\mathbf{w}_j[n]|^2 + \sigma_{b}^2},\label{eq:gamma_b_s}  
\end{equation}
respectively, where $\varpi_{rb} \in [0, 1]$  and $\varpi_{jb} \in [0, 1]$  are the residual interference levels representing the SIC quality.  
The achievable covert rates of Bob in the CCO and CCS phases can be expressed as $R_b^{\rm {\rm{cco}}}[n] = \log_2 \big(1 + \gamma_b^{\rm {\rm{cco}}}[n]\big)$ and $R_b^{\rm {\rm{ccs}}}[n] = \log_2 \big(1 + \gamma_b^{\rm {\rm{ccs}}}[n]\big)$, respectively. 

\subsection{Sensing Performance Metric}

$Q$ targets are considered that are located on the ground with the horizontal position of target $q$ being denoted as $\mathbf{s}_q = [x_q, y_q]^T$, where $q \in {\cal{Q}} \triangleq \{1,2,\cdots, Q\}$. 
In the CCS phase, Alice and Jack cooperate to enhance the covert communication performance and to collaboratively construct a hybrid monostatic-bistatic radar model. Specifically, the sensing signal in time slot $n \in {\cal{N}}_s$ comprises the dual-function signal $\mathbf{x}_a [n]$ and the AN signal $\mathbf{x}_j [n]$. Thus, the incident signal at target $q$ can be expressed as
\begin{eqnarray}
    \!\!\!\!\!\!\!\!\!y_q[n] \!&\!\!\!\!=\!\!\!&\!\! \mathbf{h}_{aq}^H \mathbf{x}_a [n] + \mathbf{h}_{jq}^H \mathbf{x}_j [n] \nonumber \\
    \!\!\!&\!\!\!\!=\!\!\!& \!\! \mathbf{h}_{aq}^H (\mathbf{w}_a [n] s_a[n] \!+\! \mathbf{x}_r[n]) \!+\! \mathbf{h}_{jq}^H \mathbf{w}_j [n] s_j[n],
\end{eqnarray}
where the first and second terms on the right-hand-side of the equality are resulted by Alice and Jack, respectively. Compared to a monostatic ISAC system, where only a single dual-function BS is deployed, the AN signal transmitted by the jamming UAV is exploited to improve the sensing performance. 
In the considered A2G-ISAC network,  Alice conducts information fusion to detect the target,  taking into account the two Tx-Rx pairs corresponding to the Alice-target-Alice and Jack-target-Alice links, respectively. Unfortunately, the Cram\'{e}r-Rao bound (CRB) of the considered A2G-ISAC network is an extremely complicated function of the hybrid monostatic-bistatic geometry, received sensing SINR, and waveforms \cite{Hybrid_Multistatic}, rendering it intractable for the design of covert communication. To address this challenge, we adopt the transmit beampattern gain as an effective metric to reflect the sensing performance \cite{Joint_Maneuver}. In particular, we consider the transmit beampattern gains from both Alice and Jack toward target $q$ and introduce a distance-normalized beampattern sum-gain as follows:
\begin{eqnarray}  
    &\!\!\!\! \!\!\!\!& \!\!\!\!\! \!\!\!\!\! \zeta(\mathbf{u}_a[n], \mathbf{u}_j[n], \mathbf{s}_q) \nonumber \\
    &\!\!\!\!=\!\!\!\!& \frac{\mathbb{E}\left[|\mathbf{a}^H(\mathbf{u}_a[n],{\mathbf{s}}_q)\mathbf{x}_a[n_s]|^2\right]}{d^2(\mathbf{u}_a[n], \mathbf{s}_q)} \!+\! \frac{\mathbb{E}\left[|\mathbf{a}^H(\mathbf{u}_j[n], \mathbf{s}_q)\mathbf{x}_j[n_s]|^2\right]}{d^2(\mathbf{u}_j[n], \mathbf{s}_q)}  \nonumber\\
    &\!\!\!\! = \!\!\!\!& \frac{\mathbf{a}^H(\mathbf{u}_a[n],{\mathbf{s}}_q)(\mathbf{w}_{a}[n_s]\mathbf{w}_{a}[n_s]^H+
        \mathbf{R}_r[n_s])\mathbf{a}(\mathbf{u}_a[n],{\mathbf{s}}_q)}{A_a^2 + \Vert \mathbf{u}_a[n] - \mathbf{s}_q \Vert^2}\nonumber\\
    &\!\!\!\! \!\!\!\!& + \frac{\mathbf{a}^H(\mathbf{u}_j[n],{\mathbf{s}}_q) \mathbf{w}_{j}[n_s]\mathbf{w}_{j}[n_s]^H \mathbf{a}(\mathbf{u}_j[n],{\mathbf{s}}_q)}{A_j^2 + \Vert \mathbf{u}_j[n] - \mathbf{s}_q \Vert^2}, \label{eq:sensing_SINR}
\end{eqnarray}
where the first and second terms on the right-hand-side of the equality are the distance-normalized beampattern gains resulted by Alice and Jack, respectively.
Signifying the power of the composite sensing signal arrived at target $q$, the introduced $\zeta(\mathbf{u}_a[n], \mathbf{u}_j[n], \mathbf{s}_q)$ 
effectively determines the detection probability or estimation error of the radar sensing \cite{Joint_Maneuver}. 

\subsection{Covertness Requirement}

Willie aims to determine whether Alice covertly transmits to Bob or not, which correspond to two hypotheses: $\mathcal{H}_0$ indicates that Alice is not sending a covert signal to Bob and $\mathcal{H}_1$ signifies that a covert transmission occurs. Under the two hypotheses, the received signals at Willie can be written as
\begin{eqnarray} 
    \mathcal{H}_0: ~y_w[n] &\!\!\!\!=\!\!\!\!& \mathbf{h}_{aw}^H\mathbf{x}_r[n]+\mathbf{h}_{{jw}}^H\mathbf{w}_{j}[n]s_j[n] + z_w[n]  \label{eq:H0}
\end{eqnarray}
and
\begin{eqnarray} 
    \mathcal{H}_1:~y_w[n] &\!\!\!\!=\!\!\!\!& \mathbf{h}_{aw}^H \mathbf{w}_{a}[n]{s}_{a}[n]
    \!+\!  \mathbf{h}_{aw}^H \mathbf{x}_r[n] \!+  \mathbf{h}_{{aw}}^H\mathbf{w}_{j}[n]{s}_{j}[n]   \nonumber\\
    &\!\!\!\! \!\!\!\!\!&  + z_w[n], \label{eq:H1}
\end{eqnarray}
respectively. In \eqref{eq:H0} and \eqref{eq:H1},
$z_w[n]\sim\mathcal{CN}(0,\sigma_w^2)$ is the AWGN at Willie.
The detection error probability (DEP) of Willie is defined as
\begin{eqnarray}
    {\xi} &\!\!\! = \!\!\!&  {\Pr} ( {{{\mathcal D}_1} | {{\mathcal H_0}}  } ) + {\Pr} ( {{{\mathcal D}_1} | {{\mathcal H_1}}  }  ),   \label{eq:DEP}
\end{eqnarray}
where ${\Pr} ( {{{\mathcal D}_1} | {{\mathcal H_0}}  }  )$ and ${\Pr} ( {{{\mathcal D}_0} | {{\mathcal H_1}}  }  )$ denote the probabilities of false alarm and missed detection, respectively, with ${\mathcal{D}_0}$ and ${\mathcal{D}_1}$ standing for Willie's binary decisions endorsing ${\mathcal{H}_0}$ and ${\mathcal{H}_1}$, respectively. 
To minimize DEP, we assume that Willie adopts the Neyman-Pearson criterion to detect the occurrence of the covert transmissions \cite{NOMA_RIS}. Thus, the optimal decision rule for Willie to minimize the DEP can be expressed as follows:
\begin{eqnarray}
    \frac{{{p_1}({y_{{w}}})}}{{{p_0}({y_{{w}}})}}\mathop  \gtrless \limits_{{\mathcal{D}_0}}^{{\mathcal{D}_1}} 1, \label{eq:likelihood_ratio}
\end{eqnarray}
where ${p_0}({y_w}) = \frac{1}{\pi \sigma _0^2} e^{ - {|y_w|^2}/{ \sigma _0^2 }} $ and ${p_1}({y_w}) = \frac{1}{\pi \sigma _1^2} e^{ -  {|y_w|^2}/{{\sigma _1^2}}} $ denote the likelihood functions of Willie's received signals under ${\cal{H}}_0$ and ${\cal{H}}_1$, respectively, with $\sigma_0^2=
\varpi_{rw}\mathbf{h}_{aw}^H\mathbf{R}_r\mathbf{h}_{aw}+\sigma_w^2$ and  $\sigma_1^2=	\vert \mathbf{h}_{aw}^H\mathbf{w}_{a} \vert^2 + \varpi_{rw}\mathbf{h}_{aw}^H\mathbf{R}_r\mathbf{h}_{aw} + \vert \mathbf{h}_{{jw}}^H\mathbf{w}_{j} \vert^2 +\sigma_w^2$, respectively. The optimal decision rule for minimizing the DEP can be rewritten as
\begin{eqnarray}
    P_w \mathop  \gtrless \limits_{{\mathcal{D}_0}}^{{\mathcal{D}_1}} {\phi ^*},\label{eq:optimal_tau}
\end{eqnarray}
where  
$ P_w \mathop = \limits^{ 
    \tilde N \to \infty} \frac{1}{\tilde N} \sum\nolimits_{\tilde n=1}^{\tilde N} |y_w(\tilde n)|^2 $ with $\tilde n$ denoting the index of the time slot and ${\phi ^ * } = \frac{{\sigma _0^2\sigma _1^2}}{{\sigma _0^2 - \sigma _1^2}}\ln \frac{{\sigma _1^2}}{{\sigma _0^2}} > 0$ 
is the optimal detection threshold that minimizes the DEP. 
According to \eqref{eq:optimal_tau}, the minimum DEP (MDEP) achieved by Willie can be derived as 
\begin{eqnarray}
    {\xi ^*} &\!\!\! = \!\!\!&    1 + {\left( {\frac{{\sigma _1^2}}{{\sigma _0^2}}} \right)^{ - \frac{{\sigma _1^2}}{{\sigma _1^2 - \sigma _0^2}}}} - {\left( {\frac{{\sigma _1^2}}{{\sigma _0^2}}} \right)^{ - \frac{{\sigma _0^2}}{{\sigma _1^2 - \sigma _0^2}}}}  . \label{eq:detection_error_probability}
\end{eqnarray}
To maintain the covertness, the MDEP needs to satisfy ${\xi^*} \ge 1 - \varepsilon$, where $\varepsilon > 0$ represents the desired covertness level for concealing the transmissions from Alice to Bob.
The MDEP involves not only the random channel realization but also the key system parameters, including the beamformers of Alice and Jack, as well as the covariance matrix of the sensing signal. For a tractable covert communication design, a lower bound on $\xi^*$ is introduced as 
\begin{eqnarray}
    {\xi ^*} &\!\!\! \ge \!\!\!& 1 - \sqrt {\frac{1}{2}{\cal D}({p_0}({y_w})||{p_1}({y_w})},~~~~ \label{eq:lower_bound}
\end{eqnarray}
where ${\cal D}({p_0}({y_w})||$ ${p_1}({y_w})) $ is the Kullback-Leibler (KL) divergence, which can be calculated as
\begin{eqnarray}
    {\cal D}({p_0}({y_w})||{p_1}({y_w})) &\!\!\! = \!\!\!& \ln \left(\frac{\sigma_1^2}{\sigma_0^2}\right) +  \frac{\sigma_0^2}{\sigma_1^2}  -1 . \label{eq:KL_divergence}
\end{eqnarray}
A stricter covertness constraint ${\mathcal D}({p_0}({y_w}) | {{p_1}({y_w})} ) \le 2{\varepsilon ^2}$ is adopted to ensure the desired covertness level. 
By substituting \eqref{eq:KL_divergence} into ${\mathcal D}({p_0}({y_w}) | {{p_1}({y_w})} ) \le\! 2{\varepsilon ^2}$, the covertness constraint ${\mathcal D}({p_0}({y_w}) | {{p_1}({y_w})} ) \le 2{\varepsilon ^2}$ can be rewritten as
\begin{eqnarray} 
&\!\!\!\!\!\!\!\!&|\mathbf{h}_{aw}^{\!H\!}\mathbf{w}_{a}|^2 +(1 \!-\!\kappa)\varpi_{rw}\mathbf{h}^{\!H\!}_{aw}\mathbf{R}_r\mathbf{h}_{aw}+  (1 \!-\! \kappa)|\mathbf{h}_{{jw}}^{\!H\!}\mathbf{w}_{j}|^2 \nonumber \\
&\!\!\!\!\!\!\!\!&~\le(\kappa \!-\! 1) \sigma^2_w, \label{eq:covert_constraint_0}
\end{eqnarray}
where $\kappa$ is the unique root of $f(\lambda ) = 2{\varepsilon ^2}$ in the interval $[1, \infty)$ and $f(\lambda ) = \ln \lambda  + \frac{1}{\lambda } - 1$ is a monotonic function with respect to $\lambda \in [1, \infty)$. 

\subsection{Optimization Problem Formulation}

The objective of the presented covert communication design is to maximize the ACR while guaranteeing the sensing performance for the A2G-ISAC network. To achieve this, an ACR maximization problem is formulated, which involves joint optimization of the dual-UAV trajectorY, dual-functional beamforming at Alice, and jamming beamforming at Jack. Specifically, we propose to optimize the key system parameters first by assuming that the entire task period is allocated as the CCO phase. Then, the dual-UAV sensing locations and corresponding beamformers are optimized for the CCS phase. 
 
In the CCO phase, the design goal is to maximize the ACR ${R}^{{\rm{cco}}}_{\rm acr} = \frac{1}{N_c}\sum_{n=1}^{N_c} R_b^{\rm cco}[n]$  by jointly optimizing the dual-UAV trajectory $\{ \mathbf{u}_a[n], \mathbf{u}_j [n]\}$,  covert communication beamformer $\{\mathbf{w}_{a}[n]\}$, and jamming beamformer $\{ \mathbf{w}_{j}[n] \}$, subject to the transmit power constraints, covertness constraint, and UAV flight constraints in \eqref{eq:position_1}, \eqref{eq:position_N}, and \eqref{eq:displacement}. The ACR maximization problem is formulated as
\begin{subequations}\label{eq:P1_sub}
    \begin{align}
        &~~(\text{P}1): \!\!\!\!\!\!\!\!\!\!\!\!\mathop{\max}\limits_{\{\mathbf{w}_{m}[n]\succeq 0, \mathbf{u}_m[n]\}}  {R}^{{\rm{cco}}}_{\rm acr}\\        ~~~~\text{s.t.}&~~~~~\Vert\mathbf{w}_a[n]\Vert^2\le P_{a}^{\max},\quad\forall n \in \mathcal{N}_c\label{eq:Pa1_max_constraint},\\
        &~~~~~\Vert\mathbf{w}_j[n]\Vert^2\le P_{j}^{\max}, \quad \forall n \in{\mathcal{N}_c},\label{eq:Pj1_max_constraint} \\ 
        &~~~~~|\mathbf{h}_{aw}^H[n]\mathbf{w}_{a}[n]|^2  
        +(1 \!-\! \kappa)|\mathbf{h}_{{jw}}^H[n]\mathbf{w}_{j}[n]|^2  \nonumber\\
        &~~~~~\le(\kappa \!-\! 1)\sigma^2_w, \forall n \in {\mathcal{N}_c}, \label{eq:covert_constraint1}\\     &~~~~~\eqref{eq:position_1},~\eqref{eq:position_N}, \text{ and } \eqref{eq:displacement}.  
    \end{align}
\end{subequations}
In problem (P1), \eqref{eq:Pa1_max_constraint} and \eqref{eq:Pj1_max_constraint} are the transmit power constraints at Alice and Jack, respectively, and \eqref{eq:covert_constraint1} is the covertness constraint in the CCO phase, which does not contain the sensing signal as that contained in \eqref{eq:covert_constraint_0}.

In the CCS phase, the two UAVs collaborate to realize both the covert communication and target sensing. The design objective is to maximize the ACR ${R}^{{\rm{ccs}}}_{\rm acr} = \frac{1}{N_s}\sum_{n=1}^{N_s} R_b^{{{\rm{ccs}}}}[n]$, 
 while ensuring compliance with the covertness and sensing requirements. The corresponding ACR maximization problem can be expressed as
\begin{subequations}\label{eq:P2_sub}
    \begin{align}
        &~~(\text{P}2): ~\mathop{\max}\limits_{\{\mathbf{w}_{m}[n], \mathbf{R}_r[n]\succeq 0, \mathbf{u}_m[n]\}} {R}^{{\rm{ccs}}}_{\rm acr}\\
        &\text{s.t.}~~\zeta(\mathbf{u}_a[n], ~\mathbf{u}_j[n], \mathbf{s}_q) 
        \ge \Gamma, \quad \forall q  \in  {\cal{Q}}, ~\forall n  \in  { \mathcal{N}_s},\label{eq:sensing_constraint}\\
        &~~~~~\Vert\mathbf{w}_a[n]\Vert^2+ {{\rm{tr}}}(\mathbf{R}_r[n])\le P_{a}^{\max},\quad\forall n \in{N_s}\label{eq:Pa_max_constraint},\\
        &~~~~~\Vert\mathbf{w}_j[n]\Vert^2\le P_{j}^{\max}, \quad \forall n \in{ \mathcal{N}_s},\label{eq:Pj_max_constraint} \\ 
        &~~~~~|\mathbf{h}_{aw}^H[n]\mathbf{w}_{a}[n]|^2  + (1 \!-\! \kappa)\varpi_{rw}\mathbf{h}^H_{aw}[n]\mathbf{R}_r[n]\mathbf{h}_{aw}[n] \nonumber\\
        &~~~~~+(1 \!-\! \kappa)|\mathbf{h}_{{jw}}^H[n]\mathbf{w}_{j}[n]|^2 \le(\kappa \!-\! 1)\sigma^2_w, ~\forall n \in  \mathcal{N}_s. \label{eq:covert_constraint}   
    \end{align}
\end{subequations}
In problem (P2), constraint \eqref{eq:sensing_constraint} represents the sensing requirement with $\Gamma$ denoting the predetermined sensing SINR threshold, \eqref{eq:Pa_max_constraint} and \eqref{eq:Pj_max_constraint} are the transmit power constraints at Alice and Jack, respectively, while \eqref{eq:covert_constraint} is the constraint on the covertness level in the CCS phase. The objective functions in problems (P1) and (P2) are non-concave due to the coupled dual-UAV trajectory and beamformers. In addition, constraints \eqref{eq:covert_constraint1}, \eqref{eq:sensing_constraint}, and \eqref{eq:covert_constraint} are non-convex due to the  complicated coupling of the design parameters. In this case, traditional convex optimization methods are insufficient to solve problems (P1) and (P2) optimally. In what follows, we sequentially tackle problems (P1) and (P2).  

\section{Covert Beamforming and Trajectory Optimization for CCO Phase}

The ACR maximization problem (P1) involves the joint optimization of the dual-UAV trajectory and dual-UAV beamforming, which remains NP-hard to solve. In this section, we decouple problem (P1) into three distinct subproblems: one for Alice's trajectory optimization, another one for Jack's trajectory optimization, and the third one for the dual-UAV covert beamforming optimization. Then, we iteratively solve the three subproblems in a BCD way to obtain a suboptimal solution for problem (P1). 

\subsection{Alice's Trajectory Optimization}

With any given covert beamformers $ \{\mathbf{w}_{a}[n], \mathbf{w}_j[n] \}$ and Jack's trajectory $ \{\mathbf{u}_j[n] \}$, problem (P1) reduces to an Alice's trajectory optimization sub-problem, which can be formulated as 
\begin{subequations}
    \begin{align}
        ~(\text{P}3):	& \mathop{\max}\limits_{\{{\mathbf{u}_a[n]}\}} ~  \frac{1}{N_c}\sum_{n=1}^{N_c}R_{{\rm{b}}}^{{\rm{cco}}}[n] \label{eq:40a}\\        \!\!\!\!\text{s.t.}&~\frac{\beta\vert\mathbf{a}^H(\mathbf{u}_a[n],\mathbf{v}_b[n])\mathbf{w}_a[n]\vert^2}{A_a^2+\Vert \mathbf{u}_a[n]-\mathbf{v}_w\Vert^2} + (1-\kappa)  \nonumber \\ 
        &\times \frac{\beta\vert\mathbf{a}^H( \mathbf{u}_j[n],\mathbf{v}_w[n] )\mathbf{w}_j[n]\vert^2}{{A_j^2+\Vert \mathbf{u}_j[n]-\mathbf{v}_w\Vert^2}}  \le(\kappa-1)\sigma^2_w, \forall n\in{\mathcal{N}_c}, \label{eq:u_a_covert_constraint2}\\
        &~\eqref{eq:position_1},~\eqref{eq:position_N}, \text{ and } \eqref{eq:displacement} .
    \end{align}
\end{subequations}
Alice's trajectory is embedded in the steering vector in problem (P$3$) in a complicated way, which results in the non-concave objective function and non-convex  covertness constraint. To obtain the optimal trajectory $\{\mathbf{u}_a[n]\}$ for problem (P$3$), we introduce the variable $\mathbf{A} (\mathbf{u}_m,  \mathbf{v}_{\bar{m}}) = \mathbf{a}^H(\mathbf{u}_m, \mathbf{v}_{\bar m}) \mathbf{a}(\mathbf{u}_m, \mathbf{v}_{\bar m}) $ with  $m \in \{a, j\}$ and ${\bar m} \in \{b, w\}$. Then, we rewrite the achievable covert rate $R^{{\rm{cco}}}_b[n]$ in a new form as
\begin{figure*}[!t]
    \normalsize
    \begin{eqnarray}
    \setcounter{equation}{36} 
   &\!\!\!\! \!\!\!\!& \!\!\!\!{\eta_{aw} \big(\mathbf{W}_a[n],d(A_a,\!\mathbf{u}_a[n],\mathbf{v}_w)\big)}     \le \bigg(\frac{{{\rm{tr}}}\big(\mathbf{W}_j[n]\mathbf{A}(\mathbf{u}_j[n],{\mathbf{v}}_w)\big)}{{d^2(A_j,\mathbf{u}_j[n],\mathbf{v}_w)}} + \frac{ \sigma_w^2}{\beta}\bigg)(\kappa -1) d^2(A_a,\mathbf{u}_a[n],\mathbf{v}_w),~\forall n  \in {{\mathcal{N}_c}}.  \label{eq:Covertness_Constraint} 
    \end{eqnarray}
     \begin{eqnarray}
    \setcounter{equation}{42} 
        \!\!\! \!\!\!\!\!& &\mu_{aw}^{(t_1)}[n]  +\big(\bm{\vartheta}_{aw}^{(t_1)}\big)^H \big(\mathbf{u}_a[n]-\mathbf{u}_a^{(t_1)}[n]\big)  
         \le  (\kappa-1)\bigg(\frac{{{\rm{tr}}}\big(\mathbf{W}_j[n]\mathbf{A}(\mathbf{u}_j[n],{\mathbf{v}}_w)\big)}{{d^2(A_j,\mathbf{u}_j[n],\mathbf{v}_w)}}+\frac{ \sigma_w^2}{\beta}\bigg)
        A_a^2 ,~~\forall n \in\mathcal{N}_c.\label{eq:Covertness_Constraint_Ua_tylor}
    \end{eqnarray}
    \hrulefill
    \vspace*{4pt}
\end{figure*} 
\begin{eqnarray} 
\setcounter{equation}{30} 
    \!{\hat{R}}^{{\rm{cco}}}_{b}[n]&\!\!\!\!\!=\!\!\!\!\!&\log_2 \! \Bigg(\!\frac{{{\rm{tr}}}\big(\mathbf{W}_a[n]\mathbf{A}(\mathbf{u}_a,\mathbf{v}_b\!)\!\big)}{{A_a^2\!+\!||\mathbf{u}_a[n]\!-\!\mathbf{v}_b||^2}}
\!\!+\!\!\frac{\varpi_{jb}{{\rm{tr}}}\big(\mathbf{W}_j[n]\mathbf{A}(\!\mathbf{u}_j,\mathbf{v}_b\!)\!\big)}{{A_j^2\!+\!||\mathbf{u}_j[n]\!-\!\mathbf{v}_b||^2}}\nonumber\\
 \!\!\!\!\!\!\! &\!\!\!\!\! \!\!\!\!\!&\!\!\!\!\!\!\!\!\!\!+\frac{\sigma_b^2}{\beta}\Bigg) \!-\!\log_2\Bigg(\frac{\varpi_{jb}{{\rm{tr}}}\big(\mathbf{W}_j[n]\mathbf{A}(\mathbf{u}_j,\mathbf{v}_b)\big)}{{A_j^2+||\mathbf{u}_j[n]-\mathbf{v}_b||^2}} + \frac{\sigma_b^2}{\beta} \Bigg) .  \label{eq:Rb_trajectory_trace_a}
\end{eqnarray} 
where $\mathbf{W}_{m}[n]= \mathbf{w}_{m}[n]\mathbf{w}_{m}^H[n]$ with $\mathbf{W}_{m}[n]\succeq0$ and ${\rm{rank}}(\mathbf{W}_m [n]) \le 1$. Now, problem (P$3$) can be equivalently rewritten as
\begin{subequations}
    \begin{align}
       (\text{P}4):	& \mathop{\max}\limits_{\{{\mathbf{u}_a[n]} \}} ~ \frac{1}{N_c}\sum_{n=1}^{N_c}\hat{R}_{b}^{{\rm{cco}}}[n]  \label{eq:uav_R_b}\\
       \!\! \text{s.t.}~
        & \frac{{{\rm{tr}}}\big(\mathbf{W}_a[n]\mathbf{A}(\mathbf{u}_a,\mathbf{v}_w)\!\big)}{A_a^2+||\mathbf{u}_a[n]-\mathbf{v}_w||^2} \!+ \!\frac{(1-\kappa){{\rm{tr}}}\big(\mathbf{W}_j[n]\mathbf{A}(\mathbf{u}_j,\mathbf{v}_w)\!\big)}{A_j^2+||\mathbf{u}_j[n]-\mathbf{v}_w||^2}  \nonumber\\ 
        &\le \frac{(\kappa-1)\sigma^2_w}{\beta},~~\forall n\in\mathcal{N}_c,\label{eq:u_a_covert_constraint_P4}\\
    &\eqref{eq:position_1},~\eqref{eq:position_N}, \text{ and } \eqref{eq:displacement}  . \label{eq:displacement_constraint}
    \end{align}
\end{subequations}	
However, in problem (P$4$), the trajectory $\{\mathbf{u}_a[n]\}$ and steering vector $\mathbf{a}(\mathbf{u}_a[n],\mathbf{v}_{\bar{m}})$ ($\bar{m} \in \{b, w \}$) are coupled in $\mathbf{A}(\mathbf{u}_a, \mathbf{v}_{\bar m})$, which exists in the objective function and the non-convex constraint \eqref{eq:u_a_covert_constraint_P4}. Thus, it is still a challenge to obtain the optimal solution for problem (P$4$). Next, we reformulated the coupled objective function and constraint into a more tractable from.  Then, we apply the SCA technique, combing with a trust-region approach, to ensure the approximation accuracy,  obtaining a suboptimal solution for problem (P$4$). 

First, we rewrite the achievable covert rate $\hat R_b^{{\rm{cco}}}[n]$ as
\begin{eqnarray}
{\hat{R}}_{b}^{{\rm{cco}}}[n] &\!\!\!\!= \!\!\!\!& \log_2 \zeta_2 [n]  - \log_2 d^2(A_a,\!\mathbf{u}_a[n] ,\!\mathbf{v}_b) \zeta_1[n] , ~~~~\label{eq:Rb_trajectory_a2}
\end{eqnarray}
where
\begin{eqnarray}
 \zeta_1[n]  &\!\!\!\! = \!\!\!\!&   \frac{\varpi_{jb}{{{\rm{tr}}}\big(\mathbf{W}_j[n]\mathbf{A}(\mathbf{u}_j[n],{\mathbf{v}}_b)\big)}}{{d^2(A_j,\!\mathbf{u}_j[n],\!\mathbf{v}_b)}}+\frac{\sigma_b^2}{\beta} , \label{eq:zeta1}
\end{eqnarray}
\begin{eqnarray}
    \zeta_2[n]  &\!\!\!\!\!=\!\!\!\!\!&  \eta_{ab}\big(\mathbf{W}_a[n],d(A_a,\mathbf{u}_a[n],\mathbf{v}_b)\big)  \!+\!  d^2(A_a,\!\mathbf{u}_a[n],\!\mathbf{v}_b) \zeta_1[n] , \!\! \nonumber \\
    &\!\!\!\!\! \!\!\!\!\!&
\end{eqnarray} 
and
\begin{eqnarray}	
&\!\!\!\!\!\!\!\!&\!\!\!\!\!\!\!\!\!\!\!\!\!\!\! \eta_{m\Bar{m}}\big(\mathbf{W}_m[n],d(A_m,\mathbf{u}_m[n],\mathbf{v}_{\Bar{m}})\big)\nonumber \\
    &\!\!\!\!=\!\!\!\!&   \sum_{\ell=1}^M\big[\mathbf{W}_m[n]\big]_{\ell,\ell}+2\sum_{k=1}^{M}\sum_{\ell=k+1}^{M}\big\vert\big[\mathbf{W}_m[n]\big]_{k,\ell} \big\vert \nonumber\\		
    &\!\!\!\! \!\!\!\!&\! \times\cos\bigg(\theta^{\mathbf{W}_m}_{k,\ell}[n]+ \frac{2\pi d_m A_m (\ell-k)}{\lambda d(A_m,\mathbf{u}_m[n],\mathbf{v}_{\Bar{m}})} \bigg) \label{eq:eta_a}
\end{eqnarray}
with $m \in \{a, j\}$ and $\bar m \in \{b, w\}$. 
In \eqref{eq:eta_a}, $\big|\big[\mathbf{W}_a[n]\big]_{k,\ell}\big|$ and $\theta^{\mathbf{W}_a}_{k,\ell}[n]$ are the magnitude and phase of $\big[\mathbf{W}_a[n]\big]_{k,\ell}$, respectively.
The derivation steps of \eqref{eq:eta_a} can be referred to \cite{Joint_Maneuver}.

To deal with the non-convex covertness constraint in problem (P$4$), the covertness constraint \eqref{eq:u_a_covert_constraint_P4} can be rewritten as \eqref{eq:Covertness_Constraint}  at the top of next page.  
However, the derived objective function and covertness constraint are still sophisticated containing the coupled trajectory $ \{\mathbf{u}_a[n] \}$, such that sub-problem of Alice's trajectory optimization cannot be solved so far. 
Next, a trust-region SCA method is proposed to optimize Alice's trajectory. Specifically, SCA is employed to address the complicated objective function and constraints, while a series of trust-region constraints are adopted in SCA to ensure the precision of the approximations.

To address the non-concave objective function ${\hat{R}}_{b}^{{\rm{cco}}}[n]$ in \eqref{eq:Rb_trajectory_a2}, in iteration $t_1$ of the trust-region SCA method, the first-order Taylor (FOT) expansion of ${\hat{R}}_{b}^{{\rm{cco}}}[n]$ is adopted as
\begin{eqnarray}
\setcounter{equation}{37} 
    {\hat{R}}^{{\rm{cco}}}_{b} [n] 
    &\!\!\!\!\!\!\!&\ge\alpha_a^{(t_1)}[n]\!+\!\big({\bm{\rho}}^{(t_1)\!}_a[n]\big)^{H}\big(\mathbf{u}_a[n]-\mathbf{u}_a^{(t_1)}[n]\big) ~~~~\nonumber\\
&\!\!\!\!\!\!\!&\triangleq{\tilde{R}}_{b}^{{\rm{cco}},(t_1)}[n], \label{eq:u_a_R_b}
\end{eqnarray}
where
\begin{eqnarray}
    \alpha_a^{(t_1)}[n] &\!\!\!\! = \!\!\!\!& \log_2 \zeta_2^{(t_1)}[n]
     -  \log_2  d^2(A_a,\!\mathbf{u}_a^{(t_1)}[n], \!\mathbf{v}_b)  \zeta_1[n] ,~~~~  
\end{eqnarray}
and
\begin{eqnarray}
    {\bm{\rho}}_a^{(t_1)} [n] &\!\!\!\! = \!\!\!\!& \frac{\bm{\gamma}_{ab}^{(t_1)}\big(\mathbf{W}_a[n],  d(A_a,\mathbf{u}^{(t_1)}_a[n],\mathbf{v}_b)\big)}{\ln{2}~ \zeta^{(t_1)}_2[n] }  \nonumber\\
    &\!\!\!\! \!\!\!\!&  + \frac{\zeta_1[n] 
      \big( \mathbf{u}^{(t_1)}_a[n] - \mathbf{v}_b \big)}{\ln{2}}  \nonumber\\
&\!\!\!\! \!\!\!\!&  \times  \Bigg( \frac{1}{\zeta^{(t_1)}_2[n]} - \frac{1}{d^2(A_a, \mathbf{u}_a^{(t_1)}[n], \mathbf{v}_b) \zeta_1[n]}  \Bigg),
~~~~~
\end{eqnarray}
with
\begin{eqnarray}
    \zeta_2^{(t_1)}[n] &\!\!\!\!=\!\!\!\!& \eta_{ab}\big(\mathbf{W}_a[n],  d(A_a,\mathbf{u}^{(t_1)}_a[n],\mathbf{v}_b) \big) \nonumber \\
    &\!\!\!\! \!\!\!\!& + d^2(A_a, \mathbf{u}_a^{(t_1)}[n], \mathbf{v}_b) \zeta_1[n]
\end{eqnarray}
and
\begin{eqnarray}
    &\!\!\!\! \!\!\!\!&\!\!\!\!\!\!\!\!
    \bm{\gamma}^{(t_1)}_{a\Bar{m}}\big(\mathbf{W}_a[n],d(A_a,\mathbf{u}_a^{(t_1)}[n],\mathbf{v}_{\Bar{m}})\big)\nonumber\\
    &\!\!\!\! = \!\!\!\!& \sum_{k=1}^{M}\sum_{\ell=k+1}^{M}4\pi\vert[\mathbf{W}_a[n]]_{k, \ell}\vert \sin\Bigg(\theta^{\mathbf{W}_a}_{k, \ell}[n] + \frac{2\pi d_a}{\lambda} \nonumber\\
    &\!\!\!\! \!\!\!\!& \times \frac{A_a (\ell - k)}{ d(A_a,\mathbf{u}_a^{(t_1)}[n],\mathbf{v}_{\Bar{m}})} \!\Bigg)  \frac{d_a A_a (\ell - k)\big(\mathbf{u}^{(t_1)}_a[n]-\mathbf{v}_{\Bar{m}}\big)}{\lambda d^3(A_a,\mathbf{u}_a^{(t_1)}[n],\mathbf{v}_{\Bar{m}})} . ~~~~~
\end{eqnarray} 

\begin{figure*}[!t]
\normalsize
\begin{eqnarray}  
\setcounter{equation}{56} 
    &\!\!\! \!\!\!& (1 - \kappa)\eta_{jw}\big(\mathbf{W}_j[n],d(A_j,\!\mathbf{u}_j[n],\!\mathbf{v}_w)\big)
\le  \Bigg( (\kappa - 1) \frac{ \sigma_w^2}{\beta}-\!\frac{{{\rm{tr}}}\big(\mathbf{W}_a[n]\mathbf{A}(\mathbf{u}_a[n],{\mathbf{v}}_w\!)\big)}{d^2(A_a,\!\mathbf{u}_a[n],\!\mathbf{v}_w)}\Bigg)d^2(A_j,\!\mathbf{u}_j[n],\!\mathbf{v}_w), ~~\forall n \!\in\! \mathcal{N}_c. \!\!\!\!  \label{eq:Covertness_Constraint_j} 
\end{eqnarray}
\begin{eqnarray} 
        \mu_{jw}^{(t_2)}[n] + \big(\bm{\vartheta}_{jw}^{(t_2)}\big)^H \!\big(\mathbf{u}_j[n]-\mathbf{u}_j^{(t_2)}[n]\big)
\le \Bigg( (\kappa-1)\! \frac{ \sigma_w^2}{\beta}-\!\frac{{{\rm{tr}}}\big(\mathbf{W}_a[n]\mathbf{A}(\mathbf{u}_a[n],{\mathbf{v}}_w)\big)}
{d^2(A_a,\!\mathbf{u}_a[n],\!\mathbf{v}_w)} \Bigg) A_j^2,~~\forall n \!\in\! \mathcal{N}_c. \label{eq:Covertness_Constraint_Taylor_j}
    \end{eqnarray}
    \hrulefill
    \vspace*{4pt}
\end{figure*} 

To tackle the non-convex covertness constraint \eqref{eq:u_a_covert_constraint_P4}, the both sides of \eqref{eq:u_a_covert_constraint_P4} are approximated by using the FOT expansion
at the local point $\big\{\mathbf{u}_a^{(t_1)}[n]\big\}$ and the approximated  covertness constraint is written as \eqref{eq:Covertness_Constraint_Ua_tylor} at the top of this page. In \eqref{eq:Covertness_Constraint_Ua_tylor},  
\begin{eqnarray} 
\setcounter{equation}{43} 
    \!\!\!\!\mu_{aw}^{(t_1)}[n] &\!\!\!\!=\!\!\!\!& {\eta_{aw}\big(\mathbf{W}_a[n],d(A_a, \mathbf{u}_a^{(t_1)}[n], \mathbf{v}_w)\big)}  \nonumber \\
&\!\!\!\! \!\!\!\!&+~(1-\kappa)\bigg(\frac{{{\rm{tr}}}\big(\mathbf{W}_j[n]\mathbf{A}(\mathbf{u}_j[n],{\mathbf{v}}_w)\!\big)}{{d^2(A_j,\!\mathbf{u}_j[n],\!\mathbf{v}_w)}}+\frac{ \sigma_w^2}{\beta}\bigg)\nonumber\\
&\!\!\! \!\!\!&\times~||\mathbf{u}_a^{(t_1)}[n]-\mathbf{v}_w||^2
\end{eqnarray}
and
\begin{eqnarray}
  \bm{\vartheta}_{aw}^{(t_1)}&\!\!\!\!=\!\!\!\!&\bm{\gamma}_{aw}^{(t_1)}[n]+(1-\kappa)\bm{\gamma}_{rw}^{(t_1)}[n]\nonumber\\
&\!\!\!\! \!\!\!\!&+~(1-\kappa)\bigg(\!\frac{{{\rm{tr}}}\big(\!\mathbf{W}_j[n]\mathbf{A}(\mathbf{u}_j[n],{\mathbf{v}}_w)\!\big)}{{d^2(A_j,\!\mathbf{u}_j[n],\!\mathbf{v}_w)}}\!+\!\frac{ \sigma_w^2}{\beta}\bigg)
\nonumber\\
&\!\!\!\! \!\!\!\!&\times~\big(\mathbf{u}_a^{(t_1)}[n]-\mathbf{v}_w\big).
\end{eqnarray}

In iteration $t_1$ of the trust-region SCA method, the trust-region constraints are introduced as 
\begin{eqnarray} 
    \big\Vert \mathbf{u}_a^{(t_1)}[n]-\mathbf{u}_a^{(t_1-1)}[n] \big\Vert \le \psi_a^{(t_1)} , ~\forall n\in {\mathcal{N}_c},\label{eq:u_a_trust_region}
\end{eqnarray}
which guarantee that Alice's trajectory is updated within the trust-region of radius $\psi_a^{(t_1)}$. In addition, the radius of the trust-region decreases adaptively as $\psi_a^{(t_1+1)} = c_{a}\psi_a^{(t_1)}$, where $0 < c_{a} < 1$ is the control factor to achieve the convergence.  
Then, the sub-problem of Alice's trajectory optimization in iteration $t_1$ of the trust-region SCA method is formulated as
\begin{subequations}\label{eq:uav_trajectory_opt_a}
    \begin{align}
        \!\!\!\!(\text{P4.}t_1): & \mathop{\max}\limits_{\{{\mathbf{u}_a[n]}\}}  \frac{1}{N_c}\sum_{n=1}^{N_c}{\tilde{R}}_{b}^{{\rm{cco}},(t_1)}[n]  \\
        \!\!\!\!\!\!&\!\!\!\!\!\!\!\!\!\!\!\!  \text{s.t.}~~ \eqref{eq:position_1},~\eqref{eq:position_N},~ \eqref{eq:displacement} ,~\eqref{eq:Covertness_Constraint_Ua_tylor},
        \text{ and } \eqref{eq:u_a_trust_region}.
    \end{align}
\end{subequations}
Now, problem (P4.$t_1$) can be efficiently solved by using standard convex optimization solvers. 
By iteratively updating Alice's trajectory $\big\{\mathbf{u}_a^{(t_1)}[n]\big\}$ and choosing a sufficiently small radius $\psi_a^{(t_1)}$ for the trust-region, the optimal solution for problem (P$4.t_1$) can always be obtained in a convergence way.

\subsection{Trajectory Optimization for UAV Jack}

For any given covert beamformers $ \{\mathbf{W}_{a}[n], \mathbf{W}_j[n] \}$ and Alice's trajectory $ \{\mathbf{u}_a[n] \}$, problem (P1) reduces to the sub-problem of Jack's trajectory optimization as 
\setcounter{equation}{46} 
\begin{subequations}
    \begin{align}
        \!\!(\text{P}5):	& \mathop{\max}\limits_{\{{\mathbf{u}_j[n]} \}} ~ \frac{1}{N_c}\sum_{n=1}^{N_c}\hat{R}_{b}^{{\rm{cco}}}[n]  \label{eq:uav_R_b_j}\\
        \text{s.t.}~~& \frac{{{\rm{tr}}}\big(\mathbf{W}_a[n]\mathbf{A}(\mathbf{u}_a,\mathbf{v}_w)\!\big)}{A_a^2+||\mathbf{u}_a[n]-\mathbf{v}_w||^2} 
         \!+\! \frac{(1\!-\!\kappa){{\rm{tr}}}\big(\mathbf{W}_j[n]\mathbf{A}(\mathbf{u}_j,\mathbf{v}_w)\!\big)}{A_j^2+||\mathbf{u}_j[n]-\mathbf{v}_w||^2}   \nonumber\\
         & \le \frac{(\kappa-1)\sigma^2_w}{\beta}, ~~\forall n \in \mathcal{N}_c, \label{eq:u_j_covert_constraint_P9}\\
    &\eqref{eq:position_1},~\eqref{eq:position_N}, \text{ and } \eqref{eq:displacement}  . \label{eq:displacement_constraint_j}
    \end{align}
\end{subequations}
Jack's trajectory is embedded in the steering vector, leading to a highly complex non-concave objective function and a non-convex covertness constraint. 
To tackle problem (P$5$), we rewrite the achievable covert rate $\hat R_b^{{\rm{cco}}}[n]$ as
\begin{eqnarray}
{\breve{R}}_{b}^{{\rm{cco}}}[n]&\!\!\!\!=\!\!\!\!&\log_2 \zeta_4 [n] - \log_2 \zeta_3 [n], ~~\forall n \in {\cal{N}}_c, \label{eq:Rb_trajectory_j2}
\end{eqnarray}
where 
\begin{eqnarray}	
  \zeta_3[n] &\!\!\!\!\! = \!\!\!\!\!& \varpi_{jb}{\eta_{jb}\big(\mathbf{W}_j[n],d(A_j,\mathbf{u}_j[n],\mathbf{v}_b)\big)}+\frac{\sigma_b^2 }{\beta}\nonumber\\
    &\!\!\! \!\!\!&\times d^2(A_j,\mathbf{u}_j[n],\mathbf{v}_b),
\end{eqnarray}
and
\begin{eqnarray}	
  \zeta_4 [n]  &\!\!\!\!\! = \!\!\!\!\!&  \zeta_3[n] + 
         \frac{{{\rm{tr}}}\big(\mathbf{W}_a[n]\mathbf{A}(\mathbf{u}_a[n],{\mathbf{v}}_b)\!\big)d^2(\!A_j,\!\mathbf{u}_j[n],\!\mathbf{v}_b\!)}{d^2(A_a,\!\mathbf{u}_a[n],\!\mathbf{v}_b)}
.~~~~ 
\end{eqnarray}
Then, the FOT expansion of \eqref{eq:Rb_trajectory_j2} is adopted as the lower bound for the objective function in iteration $t_2$ of the trust-region SCA method as 
\begin{eqnarray}
\setcounter{equation}{51}    
    \ddot{{{R}}}_{b}^{{\rm{cco}},(t_2)} [n]
    &\!\!\!\!  \triangleq \!\!\!\!&  \alpha_j^{(t_2)}[n]+\big({\bm{\rho}}^{(t_2)}_j[n]\big)^H\big(\mathbf{u}_j[n]-\mathbf{u}_j^{(t_2)}[n]\big) ~~~~ \nonumber\\
    &\!\!\!\! \le \!\!\!\!&
     {\breve{R}}_{b}^{{\rm{cco}}} [n] , \label{eq:u_j_R_b}
\end{eqnarray}
where 
\begin{eqnarray}
    \alpha_{j}^{(t_2)}[n] &\!\!\!\!=\!\!\!\!& \log_2 \zeta_4^{(t_2)}[n] -\log_2 \zeta_3^{(t_2)}[n]
\end{eqnarray}
and
\begin{eqnarray}
    {\bm{\rho}}_j^{(t_2)} [n]&\!\!\!\!= \!\!\!\!&\frac{\varpi_{jb}\bm{\gamma}^{(t_2)}_{jb}\big(\mathbf{W}_j[n],  d(A_j,\!\mathbf{u}^{(t_2)}_j[n],\!\mathbf{v}_b\!)\!\big)}{\ln{2} }\!\Bigg(  \frac{1}{\zeta^{\!(t_2)\!}_4[n]}  \nonumber\\
    &\!\!\!\! \!\!\!\!&  - ~\frac{1}{\zeta^{(t_2)}_3[n]}  \Bigg) + \Bigg(\frac{{{\rm{tr}}\big(\mathbf{W}_a[n]\mathbf{A}(\mathbf{u}_a[n],{\mathbf{v}}_b)\!\big)\!}}{{{\zeta^{(t)}_4[n]} ~d^2(A_a,\!\mathbf{u}_a[n],\!\mathbf{v}_b)}}+ \frac{\sigma_b^2}{\beta} ~~\nonumber\\
    &\!\!\!\! \!\!\!\!&  \times~\Bigg( \frac{1}{\zeta^{(t_2)}_4[n]} - \frac{1}{\zeta^{(t_2)}_3[n]}  \Bigg) \Bigg)  
\end{eqnarray}
with 
\begin{eqnarray}	
  \zeta_3 ^{(t_2)}[n]  &\!\!\!\! =  \!\!\!\!& \varpi_{jb}{\eta_{jb}\big(\mathbf{W}_j[n],d(A_j, \mathbf{u}_j^{(t_2)}[n],  \mathbf{v}_b)\!\big)} \nonumber \\
   &\!\!\!\! \!\!\!\!& + ~\frac{\sigma_b^2 d^2(A_j,\mathbf{u}_j^{(t_2)}[n],\mathbf{v}_b)}{\beta} ~~~~
\end{eqnarray} 
and
\begin{eqnarray}	
  \zeta_4^{(t_2)} [n]  &\!\!\!\! = \!\!\!\!&  \zeta_3^{(t_2)}[n] \nonumber \\
  &\!\!\!\!   \!\!\!\!& + ~
         \frac{{{\rm{tr}}}\big(\mathbf{W}_a[n]\mathbf{A}(\mathbf{u}_a[n],{\mathbf{v}}_b)\!\big)d^2(\!A_j,\!\mathbf{u}_j^{(t_2)}[n],\!\mathbf{v}_b\!)}{d^2(A_a,\!\mathbf{u}_a[n],\!\mathbf{v}_b)}
.~~~~ 
\end{eqnarray}
Similarly, we rewrite the covertness constraint \eqref{eq:u_j_covert_constraint_P9}  
as \eqref{eq:Covertness_Constraint_j} and further approximate  \eqref{eq:Covertness_Constraint_j} as \eqref{eq:Covertness_Constraint_Taylor_j}, as shown at the top of the this page, where 
\begin{eqnarray} 
\setcounter{equation}{58}
    \mu_{jw}^{(t_2)}[n] &\!\!\!\!=\!\!\!\!& (1 - \kappa){\eta_{jw}\big(\mathbf{W}_j[n],d(A_j,\!\mathbf{u}^{(t_2)}_j[n],\!\mathbf{v}_w) \big)} \nonumber\\
&\!\!\!\! \!\!\!\!& +~ \Bigg( (1 - \kappa) \frac{ \sigma_w^2}{\beta} + \frac{{{\rm{tr}}}\big(\mathbf{W}_a[n]\mathbf{A}(\mathbf{u}_a[n],{\mathbf{v}}_w) \big)}
{d^2(A_a,\!\mathbf{u}_a[n],\!\mathbf{v}_w)} \Bigg) ~~\nonumber\\
&\!\!\!\! \!\!\!\!& \times ~||\mathbf{u}_j^{(t_2)}[n]\!-\!\mathbf{v}_w||^2
\end{eqnarray}
and
\begin{eqnarray}
 \bm{\vartheta}_{jw}^{(t_2)} &\!\!\!\!=\!\!\!\!& (1\!-\!\kappa)\bm{\gamma}_{jw}^{(t_2)} + \Bigg( (1-\kappa) \frac{ \sigma_w^2}{\beta}\nonumber\\
&\!\!\!\! \!\!\!\!&  +~ \frac{{{\rm{tr}}}\big(\mathbf{W}_a[n]\mathbf{A}(\mathbf{u}_a[n],{\mathbf{v}}_w) \big)}
{d^2(A_a,\!\mathbf{u}_a[n],\!\mathbf{v}_w)} \Bigg)
\big(\mathbf{u}_j^{(t_2)}[n]-\mathbf{v}_w\big).~~
\end{eqnarray}
To ensure that Jack's trajectory updates in a feasible way, the trust-region constraints are introduced as
\begin{eqnarray}
    \big\Vert \mathbf{u}_j^{(t_2)}[n]-\mathbf{u}_j^{(t_2-1)}[n] \big\Vert \le \psi_j^{(t_2)} , ~~\forall n\in \mathcal{N}_c,\label{eq:u_j_trust_region}
\end{eqnarray}
where $\psi_j^{(t_2)}$ is the radius of the trust-region in iteration $t_2$. During iterations, the radius is updated as $\psi_j^{(t_2+1)} = c_j \psi_j^{(t_2)}$, where $0 < c_j < 1$ is the control factor.
With the above formulated concave objective function and convex constraints, the sub-problem of Jack's trajectory optimization can be reformulated as
\begin{subequations}\label{eq:uav_trajectory_opt_j}
    \begin{align}
        \!\!\!\!(\text{P5.}t_2): &~~ \mathop{\max}\limits_{\{{\mathbf{u}_j[n]}\}}  \frac{1}{N_c}\sum_{n=1}^{N_c}{\ddot{{R}}_{b}^{{\rm{cco}},(t_2)}}[n]  \\
        \!\!\!\!\!\!&\!\!\!\!\!\!\!\!\!\!\!\!  \text{s.t.}~~ \eqref{eq:position_1},~\eqref{eq:position_N}, \eqref{eq:displacement} ,~ \eqref{eq:Covertness_Constraint_Taylor_j}, 
        \text{ and } \eqref{eq:u_j_trust_region}.~
    \end{align}
\end{subequations}
Then, standard convex optimization tools can be applied to obtain the optimal solution for problem (P5.$t_2$). 

\subsection{Optimization of Covert Beamforming}

For given dual-UAV trajectory, the sub-problem of the covert beamforming optimization needs to tackle the transmit beamformer $\{\mathbf{w}_a [n]\}$ at Alice and the jamming beamformer $\{\mathbf{w}_j [n]\}$ at Jack. 
The sub-problem of the covert beamforming optimization can be written as
\begin{subequations}
    \begin{align}
        (\text{P}6):~ \!\!\!\!\!\!\!\!\!\!\!\!\!\!\!&~~\mathop{\max}\limits_{\{\mathbf{w}_{a}[n],\mathbf{w}_{j}[n]\succeq 0\}} \sum_{n=1}^{N_c} R_b^{\rm cco}[n]&\\		\text{s.t.}&~~\eqref{eq:Pa1_max_constraint},~\eqref{eq:Pj1_max_constraint}, \text{ and } \eqref{eq:covert_constraint1}.& 
    \end{align}    
\end{subequations}
In problem (P$6$), the solutions for the beamformers $\{\mathbf{w}_{a}[n],\mathbf{w}_{j}[n]\}$ in different time slots are independent with each other. Thus, the optimizations of the covert beamformers over $N$ time slots can be tackled in a separate manner for each time slot. The sub-problem of the covert beamforming optimization in time slot $n$, $\forall n \in \mathcal{N}_c$, can be formulated as
\begin{subequations}
    \begin{align}
        (\text{P}7):~~ \!\!\!\!\!\!\!\!\!\!\!\!\!\!\!&~~\mathop{\max}\limits_{\mathbf{w}_{a}[n]\succeq 0,\mathbf{w}_{j}[n]\succeq 0} R_{b}^{{\rm{cco}}} &\\		\text{s.t.}
        &~~\Vert\mathbf{w}_a[n]\Vert^2\le P_{a}^{\max}, \quad \forall n \in{\mathcal{N}_c},\label{eq:Pa_max_constraint2} \\
        &~~\Vert\mathbf{w}_j[n]\Vert^2\le P_{j}^{\max},\quad \forall n \in{\mathcal{N}_c},  \label{eq:Pj_max_constraint2} \\ 
        &~~|\mathbf{h}_{aw}^H[n]\mathbf{w}_{a}[n]|^2  + (1 \!-\! \kappa)|\mathbf{h}_{{jw}}^H[n]\mathbf{w}_{j}[n]|^2\nonumber\\
        &~~\le(\kappa \!-\! 1)\sigma^2_w,\quad \forall n \in{\mathcal{N}_c}.\label{eq:covert_constraint2}  
    \end{align}    
\end{subequations}
However, the objective function in problem (P7) is still non-concave due to the coupling of the beamformers $\{\mathbf{w}_{a}[n],\mathbf{w}_{j}[n]\}$, which makes it extremely difficult to obtain the optimal solution directly. Next, we apply SDR to tackle problem (P7).

By introducing $\mathbf{H}_{mb}[n] = \mathbf{h}_{mb}[n]\mathbf{h}_{mb}^H[n]$ and $\mathbf{W}_{m}[n]=\mathbf{w}_{m}[n]\mathbf{w}_{m}^H[n]$, where $\mathbf{W}_{m}[n]\succeq0$ and ${\rm{rank}}(\mathbf{W}_m [n]) \le 1$, the achievable covert rate in time slot $n$ can be rewritten as
\begin{eqnarray}
       \!\!\!\!\!&\!\!\!\! \!\!\!\!&\!\!\!\!\!\!\!\!{R}_{b}^{{\rm{cco}}}(\mathbf{W}_{a}[n],\mathbf{W}_{j}[n])\nonumber\\
     &\!\!\!\! \!\!\!\!&\!= \log_2\left(1+\frac{{{\rm{tr}}}(\mathbf{H}_{ab}[n]\mathbf{W}_{a}[n])}{\varpi_{jb}{{\rm{tr}}}(\mathbf{H}_{jb}[n]\mathbf{W}_{j}[n])+\sigma_{b}^2}\!\right).     \label{eq:hat_Rb}	
\end{eqnarray}
Then, problem (P7) can be reformulated as 
\begin{subequations}
    \begin{align}
        \!\!(\text{P}8):& \!\!\mathop{\max}\limits_{\mathbf{W}_{a}[n]\succeq 0,\mathbf{W}_{j}[n]\succeq 0}\!\!\!\!\!\! {R}_{b}^{{\rm{cco}}}(\mathbf{W}_{a}[n],\mathbf{W}_{j}[n]) \\ 
        \text{s.t.}~~~~ 
        &\!\!\!\!\!\! {{\rm{tr}}}(\mathbf{W}_{a}[n])\le P_{a}^{\max},\quad \forall n \in{\mathcal{N}_c}, \label{eq:Pa_trace_constraint} \\
        &\!\!\!\!\!\! {{\rm{tr}}}(\mathbf{W}_{j}[n])\le P_{j}^{\max}, \quad \forall n \in{\mathcal{N}_c}, \label{eq:Pj_trace_constraint} \\
        &\!\!\!\!\!\! {{\rm{tr}}}(\mathbf{H}_{aw}[n]\mathbf{W}_{a}[n])  + (1-\kappa){{\rm{tr}}}(\mathbf{H}_{{jw}}[n]\mathbf{W}_{j}[n])\nonumber\\
        &\le(\kappa-1)\sigma^2_w,~\forall n \in \mathcal{N}_c, \label{eq:covert_trace_constraint} \\
        &\!\!\!\!\!\! {\rm{rank}}(\mathbf{W}_{m}[n])\le 1, ~\forall m \in \{a, j\}. \label{eq:rank_constraint}
    \end{align}
\end{subequations}
Due to the rank-one constraint \eqref{eq:rank_constraint} and non-concave objective function ${R}_{b}^{{\rm{cco}}}(\mathbf{W}_{a}[n],\mathbf{W}_{j}[n])$, problem (P8)  remains non-convex. To solve problem (P8), SCA is applied to approximate the non-concave objective function to a concave one and an iteration approach is utilized to achieve the optimized solution. Let $\mathbf{W}_{a}^{(t_3)}[n]$ and $\mathbf{W}_{j}^{(t_3)}[n]$ denote the local points in iteration $t_3$, the achievable covert rate is lower bounded by $ R_b^{\rm{cco}} \ge \bar R_b^{{\rm{cco}},(t_3)}$, where $\bar R_b^{{\rm{cco}},(t_3)}$ is the  FOT expansion of $R_b^{\rm{cco}}$ in iteration $t_3$ with its form being given by
\begin{eqnarray}    &\!\!\!\!\!\!\!\!&\!\!\!\!\!\!\!\!\!\!\!{\bar{R}}_{b}^{{\rm{cco}},(t_3)}(\mathbf{W}_{a}[n],\mathbf{W}_{j}[n]) \nonumber\\
    &\!\!\!\! = \!\!\!\!& \log_2\Big({{\rm{tr}}}(\mathbf{H}_{ab}[n]\mathbf{W}_{a}[n])+{\varpi_{jb}{\rm{tr}}}(\mathbf{H}_{jb}[n]\mathbf{W}_{j}[n])  +\sigma_b^2 \Big)\nonumber\\
    &\!\!\!\!\!\!\!\!&  - \log_2\Big( {\varpi_{jb}{\rm{tr}}}\big(\mathbf{H}_{jb}[n]\mathbf{W}_{j}^{(t_3)}[n] \big) + \sigma_b^2 \Big)  \nonumber\\ 
    &\!\!\!\!\!\!\!\!& -
    {\varpi_{jb}{\rm{tr}}}\Big(	\mathbf{B}^{(t_3)}[n]\big(\mathbf{W}_{j}[n]-\mathbf{W}_{j}^{(t_3)}[n]\big)\Big),  
    \label{eq:FOT}
\end{eqnarray}
where
\begin{eqnarray}
    \mathbf{B}^{(t_3)}\!=\!\frac{\log_2(e)\mathbf{H}_{jb}[n]}{\varpi_{jb}{{\rm{tr}}}\big(\mathbf{H}_{jb}[n]\mathbf{W}_{j}^{(t_3)}[n]\big)
        +\sigma_{b}^2}.
\end{eqnarray}    
In iteration $t_3$ of SCA, by replacing the objective function ${{R}}_b^{{\rm{cco}}}(\mathbf{W}_{a}[n],\mathbf{W}_{j}[n]r)$ with its lower bound $\bar{{R}}_{b}^{{\rm{cco}},(t_3)}(\mathbf{W}_{a}[n],\mathbf{W}_{j}[n])$, the approximated version of problem (P8) in iteration $t_3$ can be written as
\begin{subequations}
    \begin{align}
&\!\!\!\!\!\!\!\!\!\!\!\!\!\! \!\!\!\!\!\!\!\!\!\!\!\!(\text{P}8.t_3)\!: \!\!\!\!\!\mathop{\max}\limits_{\mathbf{W}_{a}[n], \mathbf{W}_{j}[n]\succeq 0} \!\!\bar{{R}}_{{b}}^{{\rm{cco}},(t_3)}(\mathbf{W}_{a}[n],\mathbf{W}_{j}[n])\label{eq:beamforming_opt}\\
~~\text{s.t.}~~~~~&\eqref{eq:Pa_trace_constraint}~,\eqref{eq:Pj_trace_constraint},\eqref{eq:covert_trace_constraint}, \text{ and }\eqref{eq:rank_constraint}.
    \end{align}
\end{subequations}
For problem (P8.$t_3$), to address the rank-one constraint \eqref{eq:rank_constraint}, we introduce a penalty term $\frac{1}{\iota_1}\sum_{m\in\{a,j\}}(\Vert \mathbf{W}_m \Vert_*+\hat{\mathbf{W}}_m^{(t_3)})$ into the objective function, where $\iota_1$ is a penalty factor and $\hat{\mathbf{W}}_{m}^{(t_3)}$  is
 an upper bound on $-\Vert\mathbf{W}_m\Vert_2$ with its form $\hat{\mathbf{W}}_{m}^{(t_3)}= -\Vert\mathbf{W}_m\Vert_2^{(t_3)} - {\rm{tr}}\big({\mathbf{q}^{{(t_3)}}_{{{\rm{max}},m}}(\mathbf{q}^{{(t_3)}}_{{\rm{max}},m})^H \big(\mathbf{W}_m-\mathbf{W}_m^{(t_3)}\big)}\big)$, i.e., $-\Vert\mathbf{W}_m\Vert_2 \le \hat{\mathbf{W}}_m^{(t_3)}$. Here, $\mathbf{W}_m^{(t_3)}$ is the solution obtained in iteration $t_3$ and ${\mathbf{q}}^{{(t_3)}}_{{{\rm{max}},m}}$ is the eigenvector corresponding to the largest eigenvalue of $\mathbf{W}_m^{(t_3)}$. As $\iota_1\rightarrow{0}$, the exactly rank-one solution can be guaranteed by maximizing the  objective function due to the fact ${\rm{rank}}(\mathbf{W}_m) = 1$ is equivalent to $\Vert\mathbf{W}_m\Vert_* - \Vert\mathbf{W}_m\Vert_2 = 0$. In addition, the convex upper bound $\hat{\mathbf{W}}_m^{{(t_3)}}$ can be obtained by leveraging the FOT expansion at point $\mathbf{W}_m^{(t_3)}$. After adding the penalty term into the objective function, problem (P8.$t_3$) is reformulated as 
\begin{subequations}
    \begin{align}
        \!\!\!\!\!\!\!\!\!\!\!\!\!\!(\text{P}9.t_3):~&\!\!\!\!\!\!\!\!\!\mathop{\max}\limits_{\mathbf{W}_{a}[n],\mathbf{W}_{j}[n]\succeq 0}{\bar{R}}_{{b}}^{{\rm{cco}},(t_3)}(\mathbf{W}_{a}[n],\mathbf{W}_{j}[n]) \nonumber\\
     &   - \frac{1}{\iota_1}\sum_{m\in\{a,j\}}\big(\Vert \mathbf{W}_m \Vert_*+\hat{\mathbf{W}}_m^{(t_3)}\big). \label{eq:[penalty_rate]} \\     ~~~~~~~~~~\text{s.t.}~~~~~&\eqref{eq:Pa_trace_constraint},~\eqref{eq:Pj_trace_constraint},\text{ and } \eqref{eq:covert_trace_constraint}.
    \end{align}
\end{subequations}
Then, a convex solver can be applied to problem (P9.$t_3$) to obtain the optimal covert beamformers.

\subsection{Block Coordinate Descent Algorithm}

To obtain a suboptimal solution for problem (P1), a BCD algorithm is proposed to solve problems (P4$.t_1$), (P5.$t_2$), and (P9.$t_3$) iteratively, as summarized in Algorithm 1. In each iteration, for any given dual-UAV trajectory, the covert beamformers are optimized by solving problem (P9.$t_3$); Then, for any given covert beamformers and $\{{\bf{u}}_j[n]\}$, problem (P4$.t_1$) is solved to optimize Alice's trajectory; Likewise, prolbem (P5.$t_2$) is solved to optimize Jack's trajectory with the given covert beamformers and $\{{\bf{u}}_a[n]\}$. Untill the increasement of the ACR is no more than a pre-defined threshold, the iterations terminate.  
  
\begin{algorithm}[t]
\scriptsize
{    \caption{BCD Algorithm for Solving Problem (P1) }
    \begin{algorithmic}[1]
    \State Set $t_0 = 0$ and initialize \big\{${\mathbf{u}}_a^{(t_0)}[n]$\big\}  \text{~and~} \big\{${\mathbf{u}}_j^{(t_0)}[n]$\big\}.
     \State\textbf{repeat} $t_0 \leftarrow t_0+1$
     \State\indent\textbf{repeat}  $t_3\leftarrow t_3+1$ 
       \State \indent \indent Given $\mathbf{u}_a^{(t_0)}[n]$ and $\mathbf{u}_j^{(t_0)}[n]$, solve (P9.$t_3$) to  
       \Statex \indent\indent\indent obtain $\mathbf{W}_a^{(t_3)}[n]$ and $\mathbf{W}_j^{(t_3)}[n]$, $\forall n \in {\cal{N}}_c$ 
       \State\indent \textbf{Until}  ${\bar{R}}_{b}^{(t_3+1)}[n]-{\bar{R}}_{b}^{(t _1)} [n]\le\varphi_1$
       \State\indent Update ${\mathbf{W}}_a^{(t_0)}[n]  = {\mathbf{W}}_a^{(t_3)}[n] $ and $ {\mathbf{W}}_j^{(t_0)}[n] $ = 
       \Statex\indent\indent$ {\mathbf{W}}_j^{(t_3)}[n] $, $\forall n \in {\cal{N}}_c$ 
      \State\indent \textbf{repeat} $t_1\leftarrow t_1+1$
       \State\indent  \indent Given $\big\{\mathbf{W}_a^{(t_3)}[n]\big\}$, $\big\{\mathbf{W}_j^{(t_3)}[n]\big\}$, and 
       \Statex \indent \indent \indent $\big\{\mathbf{u}_{j}^{(t_0)}[n]\big\}$,  solve
       (P4.$t_1$) to obtain $\big\{\mathbf{u}_{a}^{(t_1)}[n]\big\}$ 
        \State \indent\textbf{Until} $ \frac{1}{N_c}\sum_{n=1}^{N_c}\big({\tilde{R}}_{b}^{{\rm{cco}},(t_1+1)}[n] \!-\! {\tilde{R}}_{b}^{{\rm{cco}},(t_1)}[n]\big) \le\varphi_2$
        \State\indent Update $\mathbf{u}_{a}^{(t_0)}[n]=\mathbf{u}_{a}^{(t_1)}[n]$, $\forall n \in {\cal{N}}_c$
         \State \indent\textbf{repeat} $t_2\leftarrow t_2+1$
       \State\indent \indent Given $\big\{\mathbf{W}_a^{(t_3)}[n]\big\}$, $\big\{\mathbf{W}_j^{(t_3)}[n]\big\}$, and  
       \Statex\indent\indent\indent $\big\{\mathbf{u}_{a}^{(t_0)}[n]\big\}$, solve (P5.$t_2$) to obtain $\big\{\mathbf{u}_{j}^{(t_2)}[n]\big\}$ 
        \State \indent\textbf{Until} $ \frac{1}{N_c}\sum_{n=1}^{N_c}\big({\ddot{R}}_{b}^{{\rm{cco}},(t_2+1)}[n] \!-\! {\ddot{R}}_{b}^{{\rm{cco}},(t_2)}[n]\big) \le\varphi_3$
        \State\indent Update $\mathbf{u}_{j}^{(t_0)}[n]=\mathbf{u}_{j}^{(t_2)}[n]$, $\forall n \in {\cal{N}}_c$
       
         \State \textbf{Until} $R_{\rm acr}^{{\rm{cco}},(t_0+1)}-R_{\rm acr}^{{\rm{cco}},(t_0)}\le \varphi_0$ \end{algorithmic}
}
\end{algorithm}

 The complexity of Algorithm 1 is discussed as follows. Problem (P4$.t_1$) and (P5$.t_2$) have $N$ optimization variables and $3N_c+2$ convex constraints respectively. The complexity of solving problems (P4.$t_1$) and (P5.$t_2$) are $\mathcal{O}\big( N^2(3N_c+2)^{1.5} \log(1/\varrho_2)\big)$, where $\varrho_2$ is the convergence accuracy. For problem (P9$.t_3$), it has $2N_c$ optimization variables and $N_c+3$ convex constraints. Thus, the complexity of problem $(P9.t_3)$ is 
$\mathcal{O}\big( (2N_c)^2 (N_c+3 )^{1.5} \log(1/\varrho_1)\big)$, where $\varrho_1$ is the convergence accuracy.

\section{Dual-Functional Beamforming and Trajectory Optimization for CCS Phase}

With the dual-UAV trajectory and dual-UAV beamforming  optimized for the CCO phase, the system parameters need to be further optimized for the CCS phase. Specifically, we need to choose the dual-UAV sensing locations for the CCS phase from the dual-UAV trajectory optimized for the CCO phase by assuming $N_c = N$. Then, the dual-functional beamforming need to be optimized with the given dual-UAV sensing locations. Determining the dual-UAV sensing locations is a minimum set coverage problem, which is NP-hard to solve.  
In what follows, we first formulate the dual-UAV trajectory optimization problem as a weighted distance minimization problem and propose a heuristic greedy algorithm to solve it. Then, the dual-functional beamforming optimization problem is formulated and tackled.
 
\subsection{Optimization of Dual-UAV Sensing Locations}

In the CCS phase, Alice and Jack collaborate to form a hybrid monostatic-bistatic radar system in addition to conducting the covert communication. From the perspective of the hybrid monostatic-bistatic radar, the distance-normalized beampattern sum-gain $\zeta(\mathbf{u}_a[n], \mathbf{u}_j[n], \mathbf{s}_q) $ is closely related to the distances from both Alice and Jack to target $q$, i.e., $d_{aq} [n]$ and $d_{jq} [n]$ with $q = 1, 2, \cdots, Q$. For given transmit powers of the dual-UAV, the sensing performance can be enhanced by  reducing the distances from Alice and Jack to target $q$. On the other hand, the distance between Alice and Bob, $d_{ab} [n]$, should be reduced to enhance the performance of covert communication. To balance the trade-off between the covert communication and sensing performances, the optimization of the dual-UAV sensing locations is formulated as a weighted distance minimization problem in the following.  
 
When the dual-UAV is scheduled to sense target $q$ in time slot $n$, we introduce $d_q^{\rm{ccs}}[n] $ to denote the weighted distance associated with the covert communication and sensing links, which is given by 
\begin{eqnarray}
    d_q^{\rm{ccs}}[n] =  \alpha_1  d_{ab}[n]  + \alpha_2   \big(d_{aq} [n]+d_{jq} [n]\big), \label{eq:sum_distance}
\end{eqnarray}
where $\alpha_1$ and $\alpha_2$ are the weighting factors satisfying $0 < \alpha_1 < 1$, $0 < \alpha_2 < 1$, and $\alpha_1 + \alpha_2 = 1$. In \eqref{eq:sum_distance}, the first and second terms on the right-hand-side of the equality are the weighted distances associated with the covert communication and sensing links, respectively. 
We assume that $N_t = N/Q$ time slots are allocated to sense each target. Based on the dual-UAV trajectory optimized for the CCO phase under $N_c = N$, determining the dual-UAV sensing locations in the CCS phase can be regarded as choosing appropriate time slots for the dual-UAV to conduct the ISAC task. Thus, we formulate the optimization of the dual-UAV sensing locations as the weighted distance minimization problem as 
\begin{subequations} 
    \begin{align}
(\text{P10}): &~~ \mathop{\min}\limits_{\mathcal{N}_s} \sum \limits_{n\in \mathcal{N}_s} \sum\limits_{q = 1}^Q  d_q^{\rm{ccs}}[n]~~~~~~ \\
 \!\!\!& ~~~~ \text{s.t.}~~ |{\mathcal{N}}_s | = N_s.~~
    \end{align}
\end{subequations}
Considering that (P10) is a minimum set coverage problem, we propose a heuristic greedy algorithm to obtain the sensing time slots that yield the minimum $\sum \nolimits_{n\in \mathcal{N}_s} \sum\nolimits_{q = 1}^Q d_q^{\rm{ccs}}[n]$, as presented in Algorithm 2. 
In Algorithm 2, the weighted distance set ${\cal D}_q = \{d_q^{\rm{ccs}}[1], d_q^{\rm{ccs}}[2], \cdots, d_q^{\rm{ccs}}[N]\}$ is first calculated for $q=1, 2, \cdots, Q$. To determine the time slots for the dual-UAV to sense target 1, the weighted distances $d_1^{\rm{ccs}}[n]$, $n \in {\cal{N}}$, are sorted in an ascending order. Among all the weighted distances $d_1^{\rm{ccs}}[n]$, $n \in {\cal{N}}$, the $N_t$ time slots resulting in the smallest $\sum \nolimits_{n\in \mathcal{N}_s}d_1^{\rm{ccs}}[n]$ are selected for the dual-UAV to sense target 1. In the step for the dual-UAV to choose the time slots to sense target $q+1$, $1 \le q \le Q-1$, the distances associated with the time slots for the dual-UAV to sense target $q$ are excluded from the sets ${\cal D}_{q+1},  {\cal D}_{q+2}, \cdots,  {\cal D}_{Q}$.   
Since the traversal calculation is conducted for only one time and the sorting for the weighted distances is also calculated for one time, Algorithm 2 can be implemented with a polynomial complexity. Based on the time slots obtained by greedy searching, the dual-UAV sensing locations are selected from the dual-UAV trajectory optimized for the CCO phase.

\subsection{Optimization of Dual-functional Beamforming}

Given the daul-UAV sensing locations, we design the joint optimization of the covert communication and sensing beamforming. Obviously, the dual-functional beamforming optimizations for different time slots in the CCS phase are independent with each other. Next, we take time slot $n$ in the CCS phase as an example to optimize the corresponding dual-functional beamforming. 

\begin{algorithm}[t]
\scriptsize
{    \caption{Greedy Algorithm for Determining Dual-UAV Sensing Locations}
    \begin{algorithmic}[1]
     \State  Calculate ${\cal D}_q = \{d_q^{\rm{ccs}}[1], d_q^{\rm{ccs}}[2], \cdots, d_q^{\rm{ccs}}[N]\}$ , $\forall q \in \{1, \cdots, Q \}$; Set $N_t = N/Q$ and $q = 1$  
    \State \textbf{repeat} $q \leftarrow q + 1$
    \State \indent Sort the distances in  ${\cal D}_q$ in the  ascending order  
    \State \indent Obtain the $N_t$ time slot indices regarding the $N_t$ smallest distances
    \State \indent Use the obtained $N_t$ time slot indices to form the set $\widehat{\cal D}_q$    
    \State \indent Exclude the elements of $\widehat{\cal D}_q$ from those of ${\cal D}_{q+1},  {\cal D}_{q+2}, \cdots,  {\cal D}_{Q}$
   \State \textbf{Until} Obtain $\widehat{\cal D}_1, \widehat{\cal D}_2, \cdots, \widehat{\cal D}_Q$
   \State Based on the dual-UAV trajectory optimized in the CCO phase, determine the dual-UAV sensing locations for the CCS phase using the time slot indices in $\widehat{\cal D}_1, \widehat{\cal D}_2, \cdots, \widehat{\cal D}_Q$ 
    \end{algorithmic}
}
\end{algorithm}

In time slot $n$ of the CCS phase, the achievable covert rate can be written as
\begin{align}
    {R}_{b}^{ccs}(\mathbf{W}_{a}[n],\mathbf{W}_{j}[n],\mathbf{R}_r[n])
    = \log_2\left(\!1\!+\!\frac{{{\rm{tr}}}(\mathbf{H}_{ab}[n]\mathbf{W}_{a}[n])}{G(\mathbf{W}_{j}[n], \mathbf{R}_r[n])}\!\right),    
\end{align}
where 
\begin{eqnarray}
&\!\!\!\! \!\!\!\!&\!\!\!\!\!\!\!\!\!\!\!\!\!\!G(\mathbf{W}_{j}[n], \mathbf{R}_r[n])\nonumber\\
&\!\!\!\! \!\!\!\!&\!\!\!\!=\varpi_{jb}{{\rm{tr}}}(\mathbf{H}_{jb}[n]\mathbf{W}_{j}[n])\!+\!{\varpi_{rb}{\rm{tr}}}(\mathbf{H}_{ab}[n]\mathbf{R}_r[n])\!+\!\sigma_{b}^2. 
\end{eqnarray}
Then, we reformualte problem (P2) as
\begin{subequations}
    \begin{align}
        \!\!(\text{P}10):& \!\!\!\!\!\!\!\!\!\!\!\!\!\!\!\!\!\mathop{\max}\limits_{\mathbf{W}_{a}[n],\mathbf{W}_{j}[n],\mathbf{R}_r[n]\succeq 0}\!\!\!\!\!\! {R}_{b}^{\rm{ccs}}(\mathbf{W}_{a}[n],\mathbf{W}_{j}[n],\mathbf{R}_r[n]) \\ 
        \text{s.t.}~~~~ & \!\!\!\!\!\! \frac{\mathbf{a}^H(\mathbf{u}_a[n],{\mathbf{s}}_q)(\mathbf{W}_{a}[n]+\mathbf{R}_r[n])\mathbf{a}(\mathbf{u}_a[n],{\mathbf{s}}_q)}{d^2(\mathbf{u}_a[n],\mathbf{s}_q)}
        \nonumber\\
        &\!\!\!\!\!\! + \frac{\mathbf{a}^H(\mathbf{u}_j[n],{\mathbf{s}}_q)\mathbf{W}_{j}[n] \mathbf{a}(\mathbf{u}_j[n],{\mathbf{s}}_q)}{d^2(\mathbf{u}_j[n],\mathbf{s}_q)}
        \ge \Gamma, \nonumber\\
        &~~~~~~~~~~~~~~~~~~~~~~~~~~~~~~~\forall q  \in  {\cal{Q}}, ~\forall n  \in  \mathcal{N}_s, \label{eq:sensing_SINR_constraint}\\
        &\!\!\!\!\!\! {{\rm{tr}}}(\mathbf{W}_{a}[n])+{{\rm{tr}}}(\mathbf{R}_r[n])\le P_{a}^{\max},~\forall n  \in  \mathcal{N}_s,  \label{eq:Pa_trace_constraint2} \\
        &\!\!\!\!\!\! {{\rm{tr}}}(\mathbf{W}_{j}[n])\le P_{j}^{\max},~\forall n  \in  \mathcal{N}_s,   \label{eq:Pj_trace_constraint2} \\
        &\!\!\!\!\!\! {{\rm{tr}}}(\mathbf{H}_{aw}[n]\mathbf{W}_{a}[n]) \!+(1-\kappa){{\rm{tr}}}(\mathbf{H}_{aw}[n] \mathbf{R}_r[n])  \nonumber \\
        &\!\!\!\!\!\! \!+\! (1-\kappa){{\rm{tr}}}(\mathbf{H}_{{jw}}[n]\mathbf{W}_{j}[n])\le(\kappa-1)\sigma^2_w,~\forall n \in  \mathcal{N}_s, \label{eq:covert_trace_constraint2} \\
        &\!\!\!\!\!\! {\rm{rank}}(\mathbf{W}_{m}[n])\le 1, ~\forall m \in \{a, j\}. \label{eq:rank_constraint2}
    \end{align}
\end{subequations}
Due to the rank-one constraint \eqref{eq:rank_constraint2} and non-concave objective function ${R}_{b}(\mathbf{W}_{a}[n],\mathbf{W}_{j}[n],\mathbf{R}_r[n])$, problem (P10)  remains non-convex. To solve problem (P10), SCA is applied to approximate the non-concave objective function with a concave one and an iteration approach is utilized to achieve the optimized solution. Let $\mathbf{W}_{a}^{(t_4)}[n]$, $\mathbf{W}_{j}^{(t_4)}[n]$, and $\mathbf{R}_r^{(t_4)}[n]$ denote the local points in iteration $t_4$, the objective function is lower bounded by $ R_b^{{\rm{ccs}}} \ge \bar R_b^{{\rm{ccs}},(t_4)}$, where $\bar R_b^{{\rm{ccs}},(t_4)}$ is the  FOT expansion of $R_b^{{\rm{ccs}}}$ in iteration $t_4$ with its form being given by
\begin{eqnarray}    &\!\!\!\!\!\!\!\!&\!\!\!\!\!\!\!\!\!\!\!{\bar{R}}_{b}^{{\rm{ccs}},(t_4)}(\mathbf{W}_{a}[n],\mathbf{W}_{j}[n],\mathbf{R}_r[n]) \nonumber\\
    &\!\!\!\! = \!\!\!\!& \log_2\Big({{\rm{tr}}}(\mathbf{H}_{ab}[n]\mathbf{W}_{a}[n])+{\varpi_{jb}{\rm{tr}}}(\mathbf{H}_{jb}[n]\mathbf{W}_{j}[n])  \nonumber\\
    &\!\!\!\!\!\!\!\!& + {\varpi_{rb}{\rm{tr}}}(\mathbf{H}_{ab}[n]\mathbf{R}_r[n]) +\sigma_b^2 \Big)\nonumber\\ 
    &\!\!\!\!\!\!\!\!& - \log_2\Big({\varpi_{rb}{\rm{tr}}}\big(\mathbf{H}_{ab}[n]\mathbf{R}_r^{(t_4)}[n]\big)  \nonumber\\
    &\!\!\!\!\!\!\!\!& + {\varpi_{jb}{\rm{tr}}}\big(\mathbf{H}_{jb}[n]\mathbf{W}_{j}^{(t_4)}[n] \big) + \sigma_b^2 \Big)  \nonumber\\ 
    &\!\!\!\!\!\!\!\!& -{\varpi_{rb}{\rm{tr}}}\Big(\mathbf{B}^{(t_4)}[n]\big(\mathbf{R}_r[n]-\mathbf{R}_r^{(t_4)}[n]\big)\Big) \nonumber\\
    &\!\!\!\!\!\!\!\!& -
    {\varpi_{jb}{\rm{tr}}}\Big(	\mathbf{C}^{(t_4)}[n]\big(\mathbf{W}_{j}[n]-\mathbf{W}_{j}^{(t_4)}[n]\big)\Big),  
\end{eqnarray}
where
\begin{eqnarray}
\mathbf{B}^{(t_4)}=\!\frac{\log_2(e)\mathbf{H}_{ab}[n]}{G^{(t_4)}(\mathbf{W}^{(t_4)}_{j}[n], \mathbf{R}^{(t_4)}_r[n])},
\end{eqnarray}
\begin{eqnarray}
\mathbf{C}^{(t_4)}\!=\!\frac{\log_2(e)\mathbf{H}_{jb}[n]}{G^{(t_4)}(\mathbf{W}^{(t_4)}_{j}[n], \mathbf{R}^{(t_4)}_r[n])}.
\end{eqnarray}    
Then, in iteration $t_4$ of SCA, by replacing the objective function ${{R}}_b^{\rm ccs}(\mathbf{W}_{a}[n],\mathbf{W}_{j}[n],\mathbf{R}_r[n])$ with its lower bound $\bar{{R}}_{b}^{{\rm ccs},(t_4)}(\mathbf{W}_{a}[n],\mathbf{W}_{j}[n],\mathbf{R}_r[n])$, problem (P10) can be approximated as
\begin{subequations}
    \begin{align}
         &\!\!\!\!\!\!\!\!\!\!\!\!\!\!\!\! \!\!\!\!\!\!\!\!\!(\text{P}11.t_4)\!:\!\!\!\!\!\!\!\!\!\!\!\!\!\!\!\!\!\!\mathop{\max}\limits_{\{\mathbf{W}_{a}[n],\!\mathbf{W}_{j}[n],\!\mathbf{R}_r[n]\succeq 0\}} \!\!\!\bar{{R}}_{{b}}^{{\rm{ccs}},(t_4)}(\mathbf{W}_{a}[n],\mathbf{W}_{j}[n],\mathbf{R}_r[n])\label{eq:beamforming_opt2}\\
~~\text{s.t.}~~~~~&\eqref{eq:sensing_SINR_constraint},~\eqref{eq:Pa_trace_constraint2}~,\eqref{eq:Pj_trace_constraint2},\eqref{eq:covert_trace_constraint2}, \text{ and }\eqref{eq:rank_constraint2}.
    \end{align}
\end{subequations}
In problem (P11.$t_4$), to address the rank-one constraint \eqref{eq:rank_constraint2}, we also introduce a penalty term $\frac{1}{\iota_2}\sum_{m\in\{a,j\}}(\Vert \mathbf{W}_m \Vert_* + \hat{\mathbf{W}}_m^{(t_4)})$, where $\iota_2$ is a penalty factor and  $\hat{\mathbf{W}}_m^{(t_4)} \triangleq \Vert\mathbf{W}_m\Vert_2 - {\rm{tr}}\big({{\mathbf{p}}^{{(t_4)}}_{{\rm{max}},m}({\mathbf{p}}^{{(t_4)}}_{{\rm{max}},m})^H \big(\mathbf{W}_m-\mathbf{W}_m^{(t_4)}\big)}\big)$ with  $\mathbf{W}_m^{(t_4)}$ denoting the solution obtained in iteration $t_4$ and ${\mathbf{p}}_{{\rm{max}},m}^{(t_4)}$ denoting the eigenvector corresponding to the largest eigenvalue of $\mathbf{W}_m^{(t_4)}$. After adding the penalty term into the objective function, problem (P11.$t_4$) is transformed as
\begin{subequations}
\begin{align}
(\text{P}12.t_4):~&\!\!\!\!\!\!\!\!\!\!\!\!\!\!\!\!\!\!\!\mathop{\max}\limits_{\mathbf{W}_{a}[n],\mathbf{W}_{j}[n],\mathbf{R}_r[n]\succeq 0}{\bar{R}}_{{b}}^{{\rm{ccs}},(t_4)}(\mathbf{W}_{a}[n],\mathbf{W}_{j}[n],\mathbf{R}_r[n]) \nonumber\\
    &- \frac{1}{\iota_2}\sum_{m\in\{a,j\}} \big(\Vert \mathbf{W}_m \Vert_*+\hat{\mathbf{W}}_m^{(t_4)}\big).  \\~~\text{s.t.}~~~~~&\eqref{eq:sensing_SINR_constraint},~\eqref{eq:Pa_trace_constraint2},~\eqref{eq:Pj_trace_constraint2}, {\text{~and~}}\eqref{eq:covert_trace_constraint2}.
\end{align}
\end{subequations}
Now, a convex solver can be applied to solve problem (P$12.t_4$) to obtain the suboptimal dual-functional beamforming for problem (P$2$).

\renewcommand{\thetable}{1} 
\setlength{\extrarowheight}{2pt}
\begin{table}[t]     
  \label{table1}
  \caption{Simulation Parameters}  
  \vspace{-0.2cm}
  \centering
  \scalebox{0.9}{
    \begin{tabular}{c|c}  
      \hline\hline
      Parameters & Values \\
      \hline
      Alice's initial location & $\mathbf{u}_a^I = [0,2]$ m\\ 
      \hline 
       Jack's initial location & $\mathbf{u}_j^I = [0,0]$ m \\
      \hline 
       Alice's final location & $\mathbf{u}_j^F= [100,0]$ m \\
      \hline 
       Jack's final location & $\mathbf{u}_j^F = [100,2]$ m \\      
      \hline 
      UAV flight altitudes & $A_{a} = 150$ m, $A_{j} = 120$ m \\
      \hline 
      Maximum flight speed & $V_{\text{max}} =15$ m/s \\      
      \hline 
      Time slot duriation & $\Delta_t = 0.5$ s \\
      \hline
      The size of ${\cal N}_s$ & $N_s =  8$ \\
      \hline 
      The residual interference level & $\varpi = -8$ dB\\
      \hline
      The number of the transmit antennas & $M = 4$\\
      \hline
      Alice's maximum transmit power & $P_a^{\rm{max}}=30$ dBm\\
      \hline
      Jack's maximum transmit power &  $P_j^{\rm{max}}=20$ dBm\\
      \hline 
      Path-loss at the reference distance & $\beta = -30$ dB \\
      \hline 
      Noise power & $\sigma_b^2 = \sigma_w^2 = -80$ dBm \\
      \hline\hline
    \end{tabular}
  }
\end{table}

\section{Simulation Results}

This section presents simulation results to validate the performance of proposed dual-UAV-aided scheme. Unless otherwise specified, the adopted simulation parameters are listed in Table I.  In the simulation, $Q$ = 4 targets are randomly located in the region of interest.  Two cases of ground locations are considered for Bob and Willie. In case I, Bob and Willie are located at $[35, 40]$ m and $[65, 40]$ m, respectively. In case II, Bob and Willie are located at $[50, 30]$ m and $[50, 35]$ m, respectively. For the comparison purpose, the following benchmark schemes are considered in the simulations. 

\begin{figure}[tb]
\begin{center}
\includegraphics[width=3.1in]{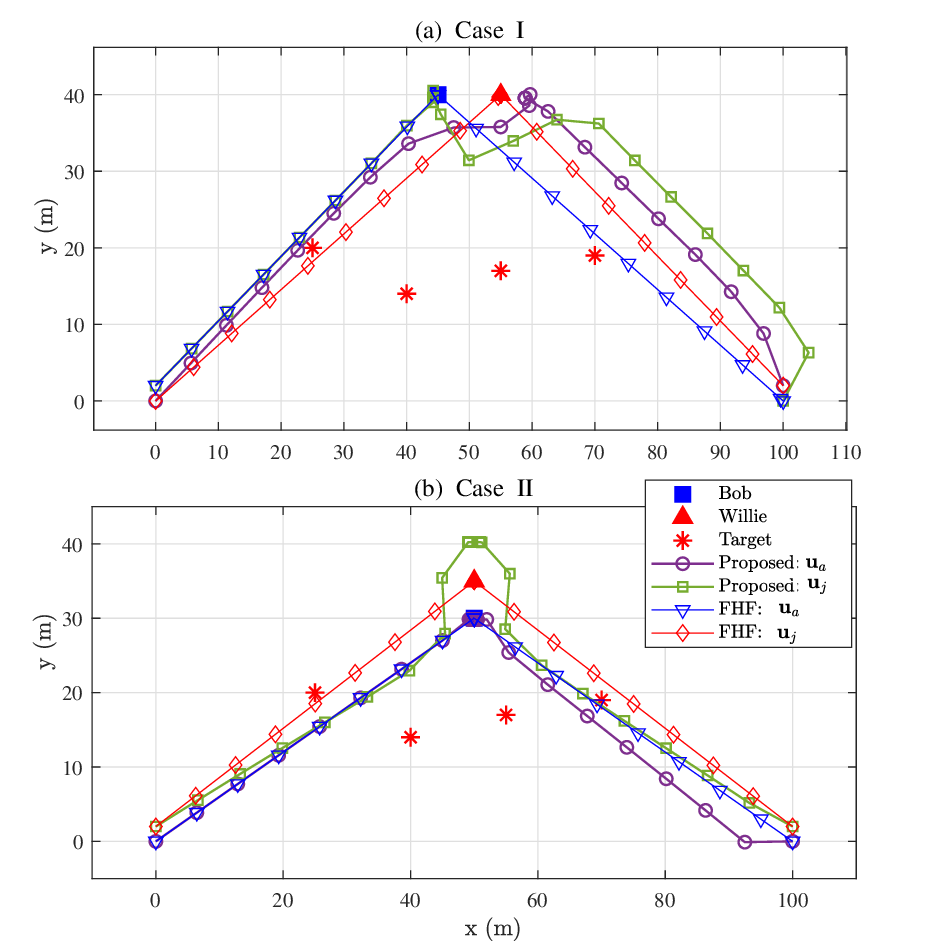}
\vspace{-0.1in}
\caption{Dual-UAV trajectories achieved by different schemes}
\end{center}
\vspace{-0.2in}
\end{figure}
 
\begin{itemize}
    \item Fly-hover-fly (FHF) + Beamforming scheme: Each UAV continuously maneuvers  along a straight line from its initial location to a ground node (Bob and Willie for Alice and Jack, respectively) at the maximum flight speed; Then, each UAV hovers on top of the ground node as long as possible; Finally, each UAV continuously maneuvers along a straight line from the ground to its final location at the maximum flight speed. The dual-UAV sensing locations and beamformers are optimized using our proposed solutions.  
    \item Dual-UAV FHF scheme: In the dual-UAV FHF scheme, the dual-UAV trajectory is conducted as in the FHF scheme, while Alice and Jack choose to hover on top of Bob and Willie, respectively. Then,  maximum ratio combing (MRC) is adopted to determine the dual-UAV beamforming with the main-lobes of the beampatterns of Alice and Jack toward Bob and Jack, respectively. 
    \item Single-UAV FHF scheme: In the single-UAV FHF scheme, only Alice is deployed using the FHF trajectory. MRC is adopted to determine Alice's beamformer with the main-lobe of the beampattern toward Bob. 
\end{itemize}

\begin{figure}[tb]
\begin{center}
\includegraphics[width=3.1in]{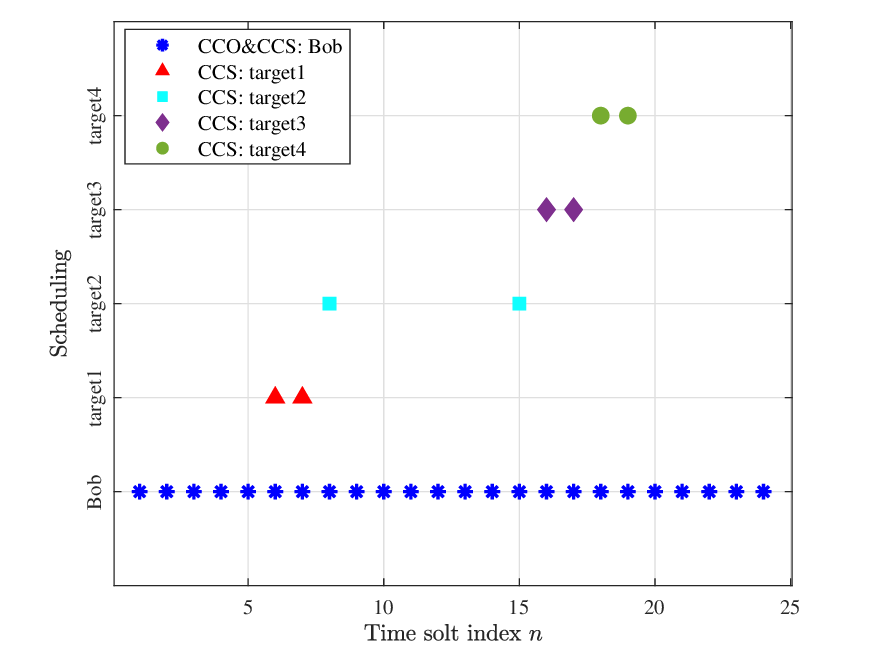}
\vspace{-0.1in}
\caption{Selected time slot indices for target sensing}
\end{center}
\vspace{-0.2in}
\end{figure}

\begin{figure}[tb]
\begin{center}
\includegraphics[width=3.2in]{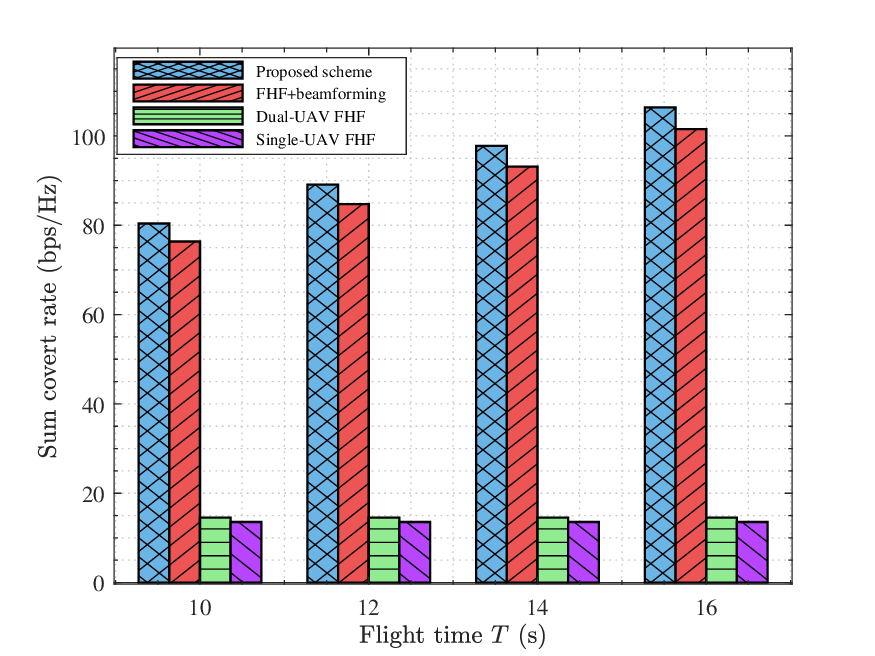}
\vspace{-0.1in}
\caption{Sum covert rate under different flight time}
\end{center}
\vspace{-0.2in}
\end{figure}

Fig. 2 illustrates the UAV trajectories obtained by different schemes in both cases I and II. No matter how the ground nodes (Bob and Willie) are located, Alice and Jack always maneuver cooperatively to confuse Willie effectively, thus improving the covert communication performance. Furthermore, the selected time slot indices in case I are presented in Fig. 3. As verified by both Figs. 2 and 3, when the dual-UAV approaches a target, the mission of the dual-UAV is to sense the target and hidden the covert transmissions simultaneously, and thus the corresponding time slots are selected for the CCS phase.  
 
The relationship between the sum covert rate and the flight time $T$ in case I is presented in Fig. 4.
As shown in Fig. 4, the proposed scheme achieves the highest sum covert rate among all the schemes. With increasing of $T$, the sum rate performance achieved by all the schemes increase.  The results in Fig. 4 also show that the sum rate performance is quite low when only the FHF schemes are applied.  

\begin{figure}[tb]
\vspace{-0.15in}
\begin{center}
\includegraphics[width=3.2in]{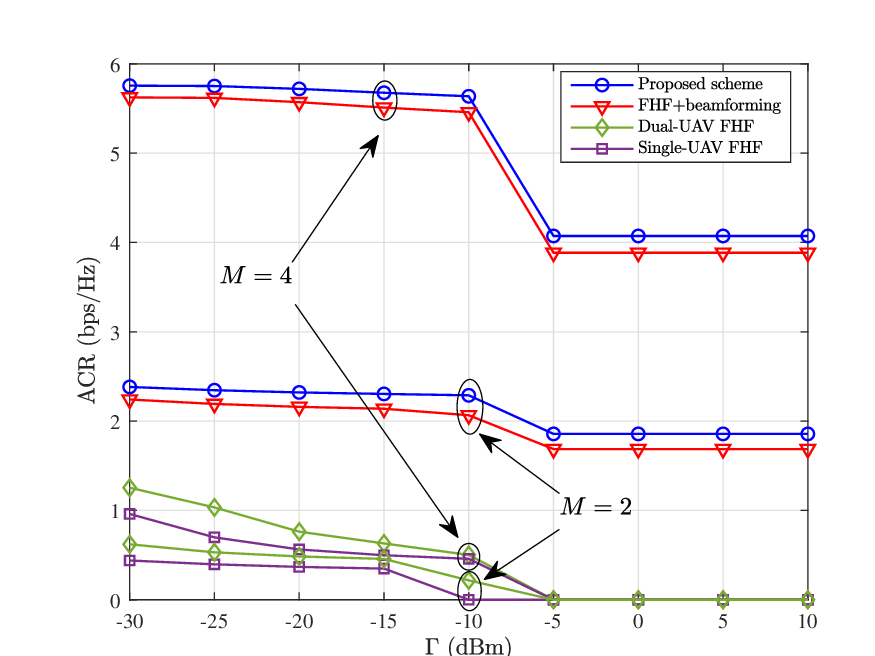}
\caption{ACR under different radar sensing thresholds}
\end{center}
\vspace{-0.1in}
\end{figure}

\begin{figure}[tb]
\vspace{-0.06in}
\begin{center} 
\vspace{0.1in}
\includegraphics[width=3.2in]{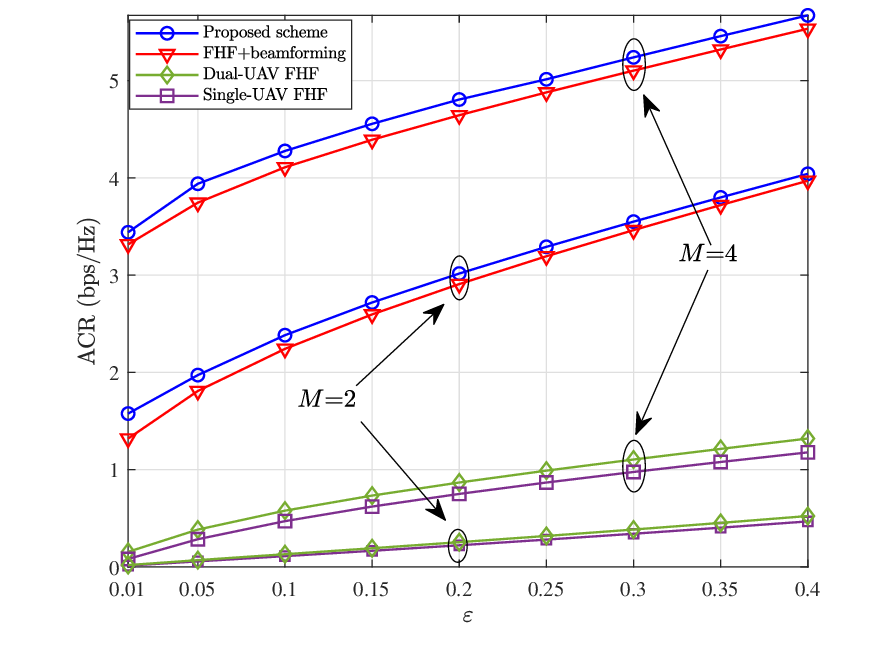}
\vspace{-0.2in}
\caption{ACR under different values of the covertness level}
\end{center}
\vspace{-0.18in}
\end{figure} 

In Fig. 5, the ACR performance under different radar sensing thresholds in case I is presented. The results in Fig. 5 show that the ACRs achieved by all the schemes decrease with increasing $\Gamma$. In particular, the dual-UAV FHF and single-UAV FHF schemes achieve the zero ACR in the high $\Gamma$ region. However, the proposed scheme achieves the highest ACR in all the considered $\Gamma$ region. In addition, with the number of the transmit antennas, $M$,  increases from $2$ to $4$, the ACRs achieved by the proposed scheme also increase. 

The impacts of the covertness level on the ACR performance in case I  are given out in Fig. 6. With increasing $\varepsilon$, the covertness constraint becomes more loose. Correspondingly, the ACRs achieved by all the schemes increase with increasing $\varepsilon$. When $\varepsilon = 0.01$, the ACRs achieved by the dual-UAV and single-UAV FHF schemes are almost zero, whereas the highest ACR is achieved by the proposed scheme. 

In Fig. 7, the effects of the residual interference level in case I are presented. Obviously, a smaller $\varpi$ leads to a higher ACR, which verifies the importance of the SIC quality on the ACR. Among all the schemes, the proposed scheme achieves the highest ACR in all the considered $\varpi$ region. In addition, the ACRs achieved by the proposed scheme and the FHF+beamforming scheme are almost same in the small $\varpi$ region. However, in the larger  $\varpi$ region, especially when  $\varpi = 0$ dB, the proposed scheme still achieves a higher ACR than the FHF+beamforming scheme. 

\begin{figure}[tb]
\begin{center}
\includegraphics[width=3.2in]{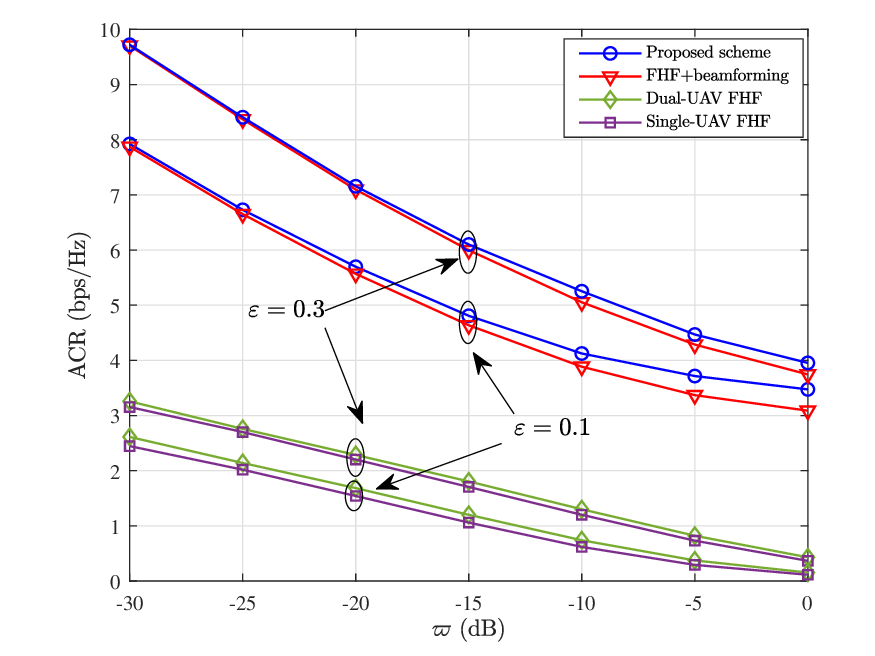}
\vspace{-0.15in}
\caption{ACR under different residual inteference levels}
\end{center}
\vspace{-0.2in}
\end{figure}

\section{Conclusions}

In this paper, we have proposed a novel dual-UAV-aided covert communication scheme to maximize the ACR by jointly optimizing the dual-UAV trajectory and dual-UAV beamforming.
It is the first time that a jamming UAV is deployed to enhance the covertness and sensing performance simultaneously for the A2G-ISAC network. Specifically, the AN signal transmitted by the jamming UAV has been  utilized not only to confuse the ground warden but also to aid the aerial BS to sense multiple ground targets by combing the target-echoed dual-functional waveform and AN components from a perspective of the hybrid monostatitc-bistatic radar. 
To maximize ACR, a BCD algorithm has been used to solve the dual-UAV trajectory and dual-UAV beamforming iteratively. A weighted distance minimization problem has been formulated to determine the dual-UAV maneuver locations suitable for sensing and a greedy heuristic algorithm has been proposed to find the corresponding solution. The simulation results have clarified the superior ACR and sensing performance achieved by the proposed scheme and revealed the effects of imperfect SIC on the system performance.  

\begin{balance}
\bibliography{IEEE_bib.bib}
\end{balance}

\end{document}